\documentclass[sigconf]{acmart}

\usepackage{calc}
\usepackage{rotating}
\usepackage{pifont}
\usepackage{subcaption}
\usepackage{multirow}
\usepackage{xspace}
\usepackage{xcolor}

\usepackage{siunitx}
\makeatletter
\renewcommand\@fnsymbol[1]{\ensuremath{\ifcase#1\or \dagger\or \ddagger\or *\else\@arabic{#1}\fi}}
\makeatother

\hypersetup{hidelinks}

\definecolor{promptbg}{gray}{0.98}
\definecolor{promptframe}{gray}{0.70}

\newcommand{\chadded}[2][]{#2}    
\newcommand{\chdeleted}[2][]{}

\DeclareRobustCommand{\ours}{\textsf{INA}\xspace}
\DeclareRobustCommand{\control}{\textsf{simple reminder}\xspace}
\DeclareRobustCommand{\baseline}{\textsf{logging only}\xspace}

\newcommand{\newdataset}{IntentionBench\xspace}

\AtBeginDocument{%
  }

\copyrightyear{2026}
\acmYear{2026}
\setcopyright{cc}
\setcctype{by}
\acmConference[CHI '26]{Proceedings of the 2026 CHI Conference on Human Factors in Computing Systems}{April 13--17, 2026}{Barcelona, Spain}
\acmBooktitle{Proceedings of the 2026 CHI Conference on Human Factors in Computing Systems (CHI '26), April 13--17, 2026, Barcelona, Spain}
\acmPrice{}
\acmDOI{10.1145/3772318.3791404}
\acmISBN{979-8-4007-2278-3/2026/04}

\begin{document}

\title{State Your Intention to Steer Your Attention: An AI Assistant for Intentional Digital Living}

\author{Juheon Choi}
\affiliation{%
  \institution{KAIST}
  \city{Seoul}
  \country{Republic of Korea}
}

\author{Juyong Lee}
\affiliation{%
  \institution{KAIST}
  \city{Seoul}
  \country{Republic of Korea}
}

\author{Jian Kim}
\affiliation{%
  \institution{Yonsei University}
  \city{Seoul}
  \country{Republic of Korea}
}

\author{ChanYoung Kim}
\orcid{0009-0008-1417-8084}
\affiliation{%
  \institution{KAIST}
  \city{Seoul}
  \country{Republic of Korea}
}

\author{Taywon Min}
\affiliation{%
  \institution{KAIST}
  \city{Seoul}
  \country{Republic of Korea}
}

\author{W. Bradley Knox}
\authornote{Equal advising.}
\affiliation{%
  \institution{Computer Science, University of Texas at Austin}
  \city{Austin}
  \state{Texas}
  \country{USA}
}

\author{Min Kyung Lee}
\authornotemark[1] 
\affiliation{%
  \institution{School of Information, University of Texas at Austin}
  \city{Austin}
  \state{Texas}
  \country{USA}
}

\author{Kimin Lee}
\authornotemark[1]
\affiliation{%
  \institution{KAIST}
  \city{Seoul}
  \country{Republic of Korea}
}

\acmSubmissionID{1128}

\begin{teaserfigure}
  \centering
  \includegraphics[width=1\textwidth]{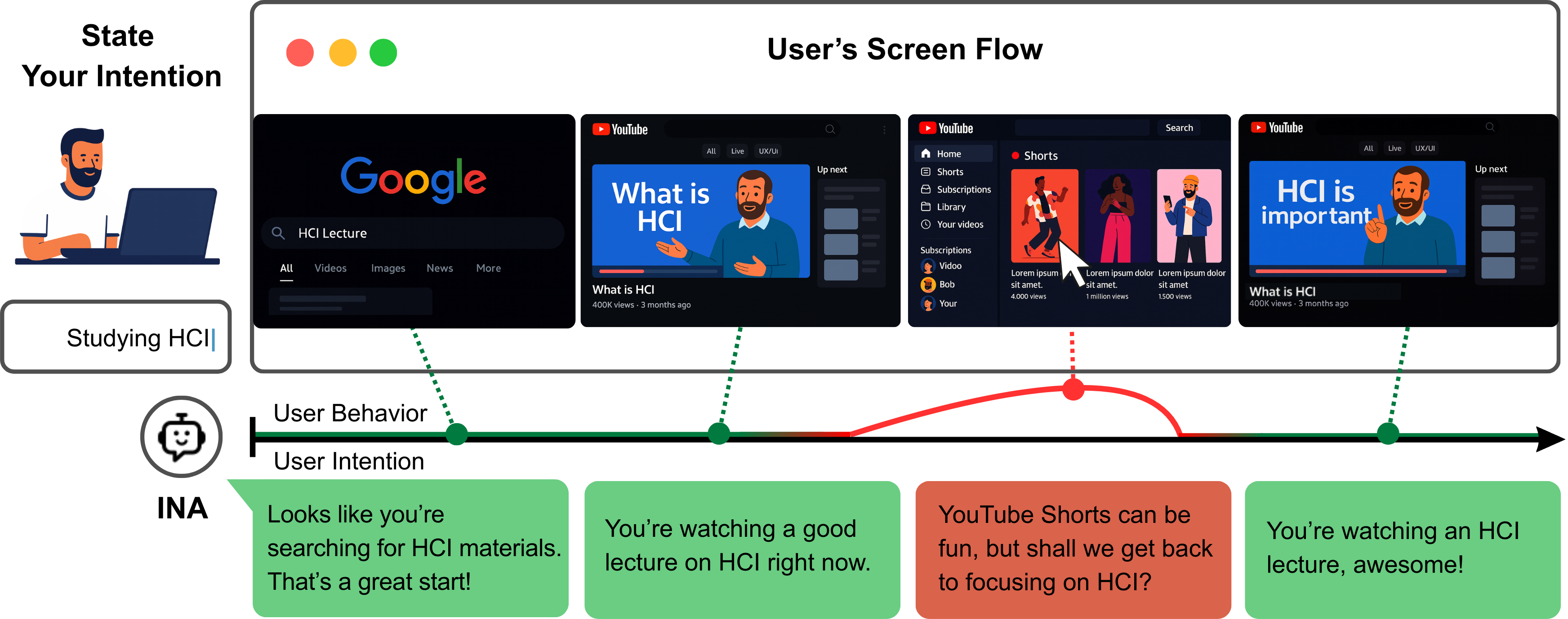}
  \Description{\textbf{Overview of Intent Assistant (\ours)}. Based on users' stated intentions, \ours monitors on-screen context to detect distractions and delivers timely interventions. By steering attention back to the stated intention, \ours supports intentional digital living. The text inside each message box is taken verbatim from what the system would actually display when operating under that intention.}
  \caption{\textbf{Overview of Intent Assistant (\ours)}. Based on users' stated intentions, \ours monitors on-screen context to detect distractions and delivers timely interventions. By steering attention back to the stated intention, \ours supports intentional digital living. The text inside each message box is taken verbatim from what the system would actually display when operating under that intention.}
  \label{fig:teaser}
\end{teaserfigure}

\begin{abstract}
When working on digital devices, people often face distractions that can lead to a decline in productivity and efficiency, as well as negative psychological and emotional impacts. To address this challenge, we introduce a novel Artificial Intelligence (AI) assistant that elicits a user's intention, assesses whether ongoing activities are in line with that intention, and provides gentle nudges when deviations occur. The system leverages a large language model to analyze screenshots, application titles, and URLs, issuing notifications when behavior diverges from the stated goal. Its detection accuracy is refined through initial clarification dialogues and continuous user feedback. In a three-week, within-subjects field deployment with 22 participants, we compared our assistant to both a rule-based intent reminder system and a passive baseline that only logged activity. Results indicate that our AI assistant effectively supports users in maintaining focus and aligning their digital behavior with their intentions. Our source code is publicly
available at \url{https://intentassistant.github.io} 
\end{abstract}



\maketitle
\section{Introduction}

Digital devices such as computers and smartphones have become essential tools for a wide range of tasks, spanning activities like studying, professional work, and planning travel. 
However, while using such devices, users are constantly assailed with tempting opportunities for distraction~\cite{Lyngs2019}, both from the device itself and from their own thoughts of alternative on-device activities they could engage in.
A variety of applications have been developed to help users stay on task~\citep{Roffarello2023, Hiniker2016, Ko2015, Kim2019, Collins2014, Ahmed2014}. 
Yet, most remain limited to rule-based blockers or simple time trackers that operate within narrow boundaries. 
Such approaches fail to capture the context of what users intend to do versus what they are actually doing, and often impose excessive restrictions when the same application serves multiple purposes.

In our formative study with eleven active computer users, we found a clear need for more intelligent, context-aware assistance beyond rule-based approaches. Participants expressed a desire for a system that could interpret their often broad and ambiguous intentions, support them in staying aligned with their goals, and provide gentle nudges when deviations occur. 
They emphasized that such an assistant should be immediate in response and gentle in manner.

In this work, we introduce the \textbf{Intent Assistant (\ours)}, an AI system that uses large language models (LLMs) to help users remain focused on their intended tasks in digital environments.
As shown in Figure~\ref{fig:teaser}, the user first enters their intention to \ours as text, which is followed by a brief clarification dialogue to capture the goal more precisely. 
\ours then continuously monitors on-screen activity, detects potential distractions, and delivers timely, gentle nudges when the user diverges from their stated intention. 
Over time, the system incorporates user feedback to refine its judgments in real time, enabling progressively more accurate and supportive interventions. 
By preserving user autonomy while gently guiding them back toward their goals, \ours serves as a collaborative assistant in everyday digital life.

We evaluate the distraction detection capability of \ours and analyze the effectiveness of our design choices.
To this end, we introduce \newdataset, a novel dataset of activity records in computer use, featuring diversity and realistic user workflows.
To create \newdataset, we collected diverse activity records by carrying out 50 unique task instructions across 14 applications and 32 websites. 
From these on-task sessions, we generated off-task samples by replacing user intentions from one session with those of another, yielding a dataset of approximately 77 hours of mixed on-task and off-task activities reflecting natural transition points (see Section~\ref{sec:dataset} for more details).
On \newdataset, \ours detects distraction from screenshot context with an accuracy of 0.878 and an F1-score of 0.845. 

Finally, we conducted a three-week in-the-wild deployment with 22 participants, comparing \ours against a \control application and a \baseline application. Participants using \ours showed a significantly lower LLM-estimated off-task ratio (0.104 vs. 0.166 with \control, $p < .001$), higher intention alignment ratings (4.44 vs. 4.23 with \control, $p < .001$), and greater focused immersion (3.74 vs. 3.34 with \control, $p = .045$; vs. 2.90 with \baseline, $p = .0003$). Beyond the quantitative outcomes, weekly surveys and interviews not only reinforced our findings but also revealed additional benefits and opportunities for improvement. Participants regarded \ours as enhancing intention fulfillment, strengthening awareness of digital habits, and providing supportive companionship. At the same time, they surfaced challenges such as notification burden, detection accuracy, and long-term user adaptation.
We summarize our main contributions as follows:
\begin{itemize}
\item Through a formative study with frequent computer users, we identify limitations of existing productivity tools and uncover users' needs for an assistant that supports intention-focused digital activity. These insights informed the design goals of our system.
\item We present the design and implementation of the Intent Assistant (\ours), an AI assistant that clarifies user intentions, detects context-dependent distractions and provides timely, non-intrusive interventions.
We develop and publicly release the \newdataset dataset, and use it to evaluate \ours’s distraction detection capability.
\item We validate \ours in a three-week in-the-wild deployment with 22 participants, demonstrating its effectiveness in reducing off-task behavior, enhancing focus, and surfacing practical challenges for further improvement of intention-aligned assistance.
\end{itemize}
Together, these contributions advance the design of context-aware, ambient digital assistants by demonstrating how AI systems can pioneer adaptive, intention-aligned collaboration in everyday computing.
\section{Related Work} 
\chadded{In this section, we review relevant areas.
These include the theoretical foundation of our work, existing digital self-control tools and their limitations, and recent AI-based systems in related domains.}

\subsection{\chadded{Intention-Behavior Gap and Dual-Systems Theory}} \label{sec:dualsystem}

\chadded{People frequently experience an intention–behavior gap~\cite{Sheeran2016}, where their actions fail to align with their goals in domains such as health, academics, and work. This gap arises not only from limited willpower but from an interplay of factors, including insufficient execution plans, habitual behaviors, and environmental influences~\cite{Grimmer2016,Casais2021,Wee2022}. Repeated failures to bridge this gap can reduce self-efficacy and increase psychological stress~\cite{Adriaanse2021,Chow2015}. 
These challenges have intensified in the digital era, as engagement-driven platforms~\cite{Lyngs2020,Lukoff2021,Du2021} and constant notifications~\cite{Pielot2017,Okoshi2017} introduce persistent distractions that make it harder for people to follow through on their intentions in everyday digital life~\cite{Meier2021,Du2021}.}

\chadded{We frame the intention–behavior gap in digital activities through Dual-Systems theory~\cite{Hofmann2009,Kahneman2003}, which distinguishes fast, automatic reactions (System 1) from slow, deliberate control (System 2)~\cite{Stanovich2000}. In digital environments, self-control failures arise when System 2 falters, allowing habitual System 1 responses (such as clicking notifications or recommendations) to override goal-directed intentions~\cite{Lyngs2019}. Because System 2 is slow and resource-limited~\cite{Miller1956,Cowan2010}, intention–behavior gaps readily emerge~\cite{Kotabe2015}. In this work, we propose a method that provides timely intention reminders~\cite{Lyngs2019}, thereby re-engaging System 2 to prevent automatic, unintentional System 1 behaviors from dominating.
}

\subsection{Digital Self-Control Tools}\label{sec:dsct}
\chadded{Digital Self-Control Tools (DSCTs) assist users in reducing the intention–behavior gap in their digital activities~\cite{Roffarello2023, Lyngs2019,Shen2019,Hiniker2016}. A representative approach involves rule-based tools that block specific applications or websites or limit usage time~\cite{Hiniker2016, Ko2015}. When users violate pre-set rules, these tools introduce friction, such as locking devices~\cite{Kim2019,Ko2016}, displaying warning messages~\cite{Hiniker2016,Tseng2019}, or requiring the typing of specific phrases~\cite{Cox2016}, to discourage undesirable behavior.
Another category of tools promotes self-reflection by visualizing usage patterns or displaying pop-ups that remind users of goals or tasks~\cite{Shen2019, Rooksby2016}.}

\chadded{Despite their prevalence, most approaches are limited because they rely on static rules defined at the application or domain level~\cite{Roffarello2023}. Such static systems fail to distinguish between different digital contexts and therefore block access uniformly, often causing frustration~\cite{Lukoff2021, Oduor2016}. 
Similarly, self-reflective pop-ups often appear as routine notifications unrelated to the user’s current context, leading users to ignore them and reducing their effectiveness~\cite{Kovacs2018, Kovacs2021}.
To address these limitations, recent work has explored leveraging Large Language Models (LLMs) to deliver more dynamic and personalized interventions~\cite{Wu2024,Li2024,Kim2024}. However, these systems typically operate through chat-based interfaces, making it difficult to infer the user’s precise context or intervene seamlessly during ongoing tasks~\cite{Roffarello2023, Okeke2018, Scibetta2025}.
In this work, we explore an AI-based, context-aware assistant that provides timely interventions based on users’ stated intentions and their on-screen activity.
}

\subsection{\chadded{AI-driven Intention Understanding}}
\chadded{Recent advances in AI have demonstrated strong capabilities in understanding multimodal data and contextual information~\cite{brown2020language,radford2021learning,alayrac2022flamingo,install2025}.
In contrast to rule-based systems that depend heavily on costly and labor-intensive human specifications, AI-based approaches provide more adaptive and fine-grained mechanisms for interpreting user behaviors and inferring underlying intentions across diverse application domains, including context-driven LLM interfaces or situated AI systems~\cite{Jaber2024,Arakawa2025,Dang2025,
fang2025mirai,jones2024goalcoach}.
For instance, \citet{Chen2023} developed a model capable of distinguishing time-killing behaviors on mobile devices with high precision.
\citet{Shaikh2025} built a general user model for intention prediction, demonstrating a level of universality based on the common knowledge and context-awareness embedded in LLMs.
These studies indicate that AI methods can function as reliable proxies for inferring user intent.
Building on this insight, in this work, we leverage AI-based semantic alignment as an indicator of behavioral deviation and introduce a novel assistant system for digital self-control.}

\section{Formative Study for Assistant Design}
\chadded{While existing literature explains the phenomenon of digital distraction well, there is a lack of concrete guidelines on when and how an AI assistant should intervene within the screen-based digital activity context. To gain design insights for concrete functionalities and interaction strategies for such an assistant,} we conducted formative interviews with individuals who use personal computers for their major activities. We describe the study procedure (Section~\ref{sec:formative}), present key findings (Section~\ref{sec:findings}), and outline the resulting design goals (Section~\ref{sec:design_goals}).

\subsection{Participants and Procedure} \label{sec:formative}

We recruited 11 participants (3 undergraduates, 4 graduate students, 4 office workers; ages 20–30, average age 25.6; 7 male, 4 female), whom we refer to as F1-F11. \chadded{Participants were recruited individually through word-of-mouth within a university community.} All participants reported using personal computers for at least four hours daily.
\chadded{We determined this group suitable for uncovering in-depth contexts and concrete needs for assistant design, considering its characteristic of being highly accustomed to digital environments.}
Eight participants had prior experience with productivity tools such as blocking or time-tracking applications, offering insights into the limitations of widely used conventional systems.

We conducted interviews focusing on three areas. 
First, we examined participants' digital intentions: the goals they set for computer use, the strategies they employed to pursue these goals, and the patterns of deviation that emerged.
Second, we investigated their experiences with existing productivity tools, including perceived benefits, limitations, and reasons for discontinuation.
Third, we asked participants to envision an intelligent, context-aware assistant that could better support intentional digital activity, focusing on desired features, intervention styles, and timing. 
All interviews were audio-recorded with prior consent, manually transcribed. \chadded{We analyzed the data using inductive thematic analysis~\cite{braun2006using}. Two researchers independently coded the transcripts and refined the themes through iterative discussion.}
Participation was voluntary and uncompensated.

\subsection{Findings} \label{sec:findings}

\paragraph{Ambiguous intentions and context-dependent distractions}
Participants were asked about their typical focus periods and how they would set intentions when carrying out tasks on digital devices, providing up to three examples. 
Intentions were often expressed in high-level terms (e.g., ``coding'', ``reading papers'', ``finish debugging''; F1, F3, F9) rather than in detailed statements. 
These broad expressions often encompassed multiple subtasks (e.g., searching for resources, implementing) and context-specific exceptions, making precise interpretation difficult.

Focus periods typically lasted 30–60 minutes but were often interrupted by distractions. 
Such distractions were commonly triggered by workflow stalls (e.g., reaching a dead end while debugging), boredom, or application notifications (e.g., email). Importantly, these patterns were highly context dependent: the same applications could serve as either task-related tools or sources of distraction depending on use. For example, YouTube enables both studying course videos and casual browsing, and email allows both on- and off-task communication.

\paragraph{Limitations of existing tools}
Seven of the eight participants who had prior experience with productivity tools expressed dissatisfaction (F1--F4, F8, F10--F11), noting that these systems rely heavily on static, rule-based methods and therefore fail to account for context. 
As a result, users often had to manually disable blocking features to perform legitimate activities (e.g., accessing YouTube to study English, F11), which exposed them to additional distractions. 
Participant F11 described this frustration:  \textit{``I had to forcibly disable the blocking program for studying purposes, only to get distracted and waste more time''}.
Such experiences underscore the limitations of static rules and point to the need for more intelligent, context-aware approaches.

\paragraph{Envisioning context-aware assistance}

Participants described what an envisioned assistant for supporting intentional activity might look like. 
They preferred an assistant that could understand the user’s intention context, adapt flexibly to different situations, and intervene promptly when unintentional deviation from the stated intention was detected (F1, F9, F11). 
Rather than rigid blocking mechanisms, which were often perceived as intrusive and compulsory, participants expressed a strong preference for notification-based interventions that respected user autonomy. 
They also emphasized the importance of adjustable controls, such as fine-tuning how frequently the assistant should intervene or how persistent notifications should be (F3, F5, F6).

Tone and style of interventions were another recurring theme. 
While most participants preferred a gentle and polite tone for notifications (F4, F7), some considered friendly or more direct wording effective depending on the situation. 
Beyond distraction management, participants also envisioned interventions that could enhance productivity, such as offering praise or encouragement (F8), assisting with planning (F10), or providing intention-related suggestions when encountering obstacles (F8, F9). 
Taken together, these perspectives portray an assistant that is immediate in response, gentle in manner, and flexible in approach, not a strict enforcer but instead collaborative.

\subsection{Design Goals} \label{sec:design_goals} 
Based on the formative study, we derive four key design goals (DGs) to guide the development of a context-aware assistant.
\begin{itemize} 

\item \textbf{DG1: Intention Understanding.}
The assistant must first enable users to articulate their intentions. Because such intentions are often abstract and open to multiple interpretations, the assistant should clarify them through interaction. Clearer intentions allow the assistant to anticipate on-task actions more effectively and minimize interventions that are based on incorrect detection of distraction.

\item \textbf{DG2: Context-Aware Distraction Detection.}
The assistant should understand the user activity flexibly, going beyond simple blacklist-based rules. To this end, it must intelligently analyze contextual cues (such as on-screen content) in relation to user intention to understand their activity continuously and adaptively.

\item \textbf{DG3: Timely and Gentle Interventions.}
When a distraction is detected, the assistant should intervene promptly to help guide the user back to their original intention. 
Interventions should be polite and dismissible notifications rather than enforced restrictions. 
The assistant may also reinforce positive behaviors by offering praise or encouragement when the user returns to their intended activity.

\item \textbf{DG4: Feedback-Driven Refinement.} 
Given that any contemporary assistant will detect distraction imperfectly, the user should be able to provide feedback on its detection to improve its future detection accuracy. This feedback can be used not only to improve general detection but also user-specific detection, for cases when the correct judgment of distraction in one context depends on the specific user. 

\end{itemize}

\newcommand{\cmark}{\ding{51}} 
\newcommand{\xmark}{\ding{55}} 

\begin{figure*}[t] 
\centering
\includegraphics[width=\textwidth]{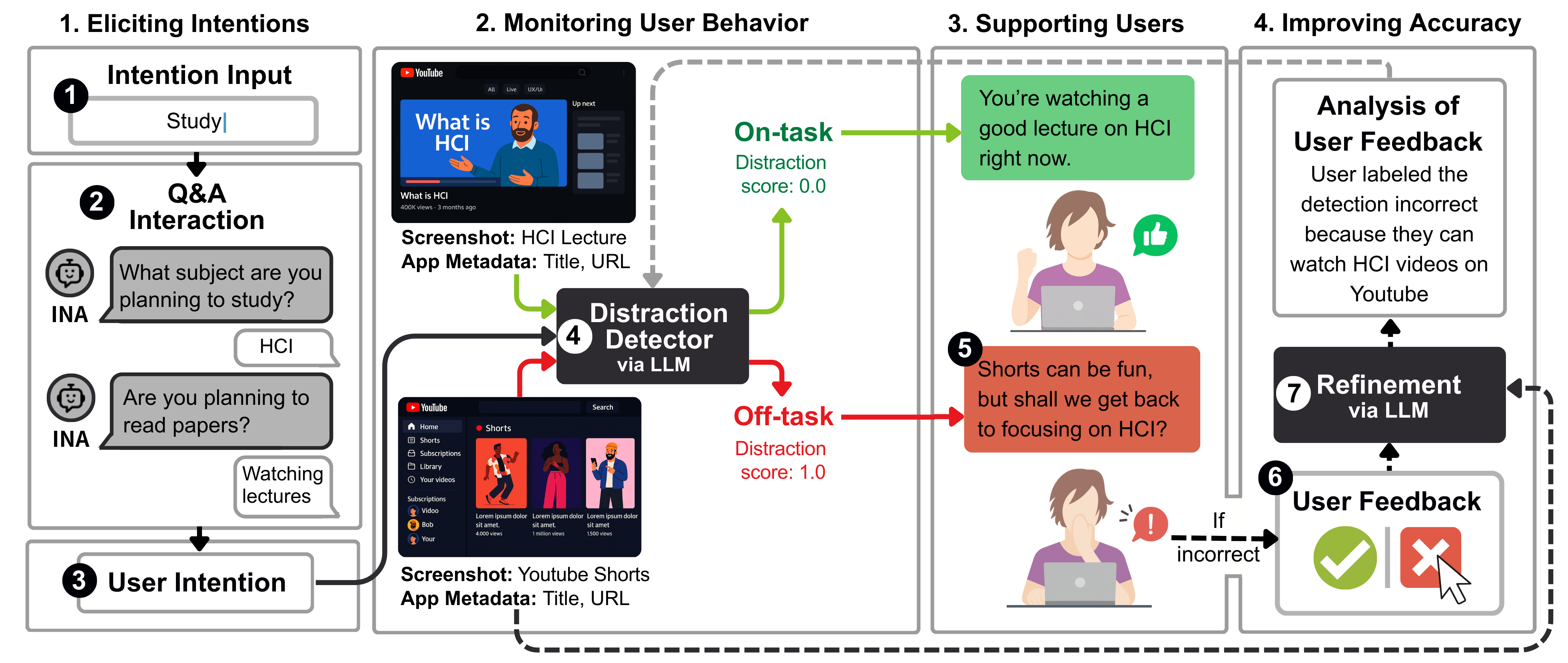}
\Description{A diagram showing the workflow of INA. When a user inputs an intention, INA elicits the intention, monitors the user behavior with a distraction detector module, provides supportive nudges, and improves accuracy through user feedback.}
\caption{Overview of \ours. 
(1) The process begins with a user's initial, often abstract, intention. (2) The system engages in an LLM Q\&A interaction to (3) establish a clarified user intention. (4) An LLM detector then analyzes on-screen activity every two seconds, using a distraction score to classify the user's state as on-task or off-task. (5) Based on this state, the system delivers a gentle nudge or positive reinforcement. (6) The user can provide feedback on this intervention, which (7) the LLM uses to refine its model, improving detection accuracy for future interactions.
}
\label{fig:system-design}
\end{figure*}
\section{Intent Assistant}
\label{sec:overview}

In this section, we introduce the \textbf{Intent Assistant (\ours)}, an AI assistant that supports users by helping them accomplish their intended screen-based tasks. 
Our key idea is to build a context-aware system around a large language model (LLM), drawing on the LLM's common-sense reasoning capabilities.
As shown in Figure~\ref{fig:system-design}, \ours consists of four main components, each designed to address the design goals (DGs) presented in Section~\ref{sec:design_goals}.
Each \textit{session} begins when the user enters their intention as text and ends when the user presses the stop button. 
From this input, \ours asks a few clarifying questions regarding their intention (DG1). 
The system continuously monitors the user’s on-screen activity to detect when they diverge from the stated intention (DG2). 
After detecting such distractions, \ours intervenes with gentle, dismissible notifications (DG3). 
Over time, the system incorporates user feedback on the correctness of its detections to improve its detection accuracy (DG4).
In the following subsections, we describe each component in detail.

\subsection{Eliciting User Intentions}\label{subsec:intention-caputring}

A session with \ours begins when the user enters their intention as text. 
However, in our pilot study, we observed that user inputs vary widely, ranging from abstract statements (e.g., ``study'') to very specific ones (e.g., ``study HCI concepts in YouTube lectures''). 
While specific inputs make it easier for the system to relate user behaviors to their intentions, abstract inputs make this process more difficult.

To address the abstract nature of user intentions, we incorporate a \textit{clarification} process with short Q\&A interaction using the LLM.
Given an initial intention, the LLM generates a clarifying question to better understand the user’s goal. 
If the user responds, the LLM may generate a follow-up question to further refine the expressed intention.
For example, when the initial input is ``study’’, the LLM may first ask ``What subject are you planning to study, such as math, history, or a language?'', narrowing the scope to a specific domain.
If the user answers, ``I will study HCI concepts’’, the LLM can then follow up with “What tools or resources will you use, such as textbooks or online courses?'' shifting the focus from what to how the user intends to study.
The LLM asks two questions and allows the user to stop the dialogue at any time.
Please refer to the Appendix~\ref{app:clarification} for the details of this procedure.

\subsection{Monitoring User Behavior to Detect Distraction}\label{subsec:monitoring}

Given a user's clarified intention from the Q\&A process, our system continuously monitors on-screen activities to evaluate whether users remain focused on their stated goals. A central challenge in this process is that distractions are not always identifiable from surface-level signals such as URLs, application names, or keywords. 
For example, two users may both be watching videos on YouTube: for one, the activity directly supports their goal of studying HCI concepts, while for the other, it constitutes an unrelated diversion.

To address this challenge, we leverage LLMs to assess the semantic alignment between user activities and their stated intentions. 
Specifically, the LLM is prompted to evaluate alignment and generate a \textit{distraction score}—a score in $[0,1]$ that ranges from perfect alignment (0) to complete misalignment (1)—based on three input signals:
\begin{itemize}
\item \textbf{Intention}: The clarified intention that was generated by the LLM from both the original intention and the Q\&A process
\item \textbf{Current screenshot}: A single screenshot of the user’s current screen
\item \textbf{Application metadata}: Information about the application in focus, including its title and, if the application is a browser, the current URL\footnote{This application metadata is especially important since certain contents (e.g., streaming videos) may not appear in screenshots because the platform prevents capturing them.}
\end{itemize}
\begin{figure*}[t]
    \centering
    \includegraphics[width=1.01\linewidth]{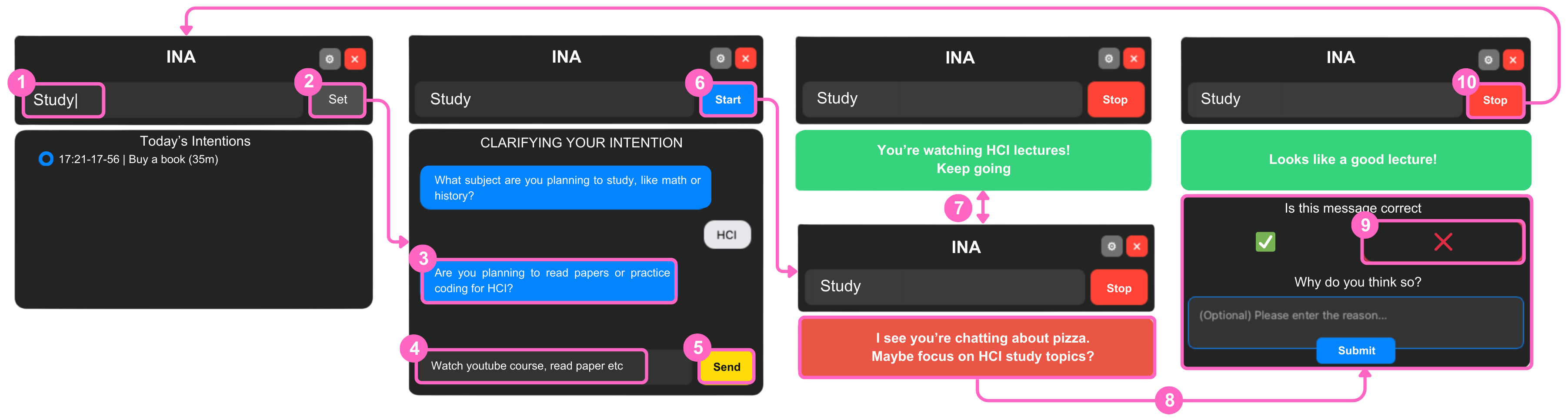}
    \caption{
    (1-2) The user inputs an intention and presses `Set' to initiate a session. (3-5) A chat interface is used for a brief Q\&A clarification process, where the user responds to the LLM's questions. (6) The user can also press `Start' to skip this clarification and begin the session immediately. (7) During the session, \ours displays the user's current state with colored notifications: green for on-task and red for off-task. (8-9) The user can hover over a notification to provide feedback, marking it as correct or incorrect and optionally adding a reason. (10) The `Stop' button can be pressed at any time to end the session.
    }
    \Description{Interface screenshot of \ours, showing how a user sets an intention input, clarifies it through Q&A, receives on-task or off-task notifications, and provides feedback during the session.}
    \label{fig:system_implementation}
\end{figure*}
By integrating these inputs, the LLM assesses how observed behavior cues (e.g., text, images, UI elements, application functions) relate to the user’s intention, rather than relying solely on keywords or application categories. 
The output format encourages the model to generate a brief chain-of-thought rationale before producing a distraction score, which improves both accuracy and interpretability (see Appendix~\ref{app:detection} for the full prompt).
The resulting distraction score provides the quantitative basis for the subsequent process, i.e., judging whether the user is distracted and delivering notifications for gentle guidance.

\subsection{Supporting Users According to their Distraction State} 
\label{subsec:interventions} 

Given a distraction score, \ours classifies the distraction state of the user as \textit{on-task} if the distraction score is less than 0.5—indicating that the user is currently aligned with their stated intention—or as \textit{off-task} otherwise, meaning they are distracted. 
Upon transition from one distraction state to another, a notification is issued if and only if a change between these states is sustained for a certain duration (4 seconds), rather than responding to brief or accidental shifts. 
Specifically upon an on-task$\rightarrow$off-task transition, \ours delivers a gentle nudge that invites the user to return to their stated intention. 
On the other hand, upon an off-task$\rightarrow$on-task transition, the system provides praise to reinforce the user's regained focus. 
If the user remains in the off-task state, the nudge is repeated at fixed intervals (every 30 seconds), a compromise chosen to remain noticeable without being disruptive, whereas no further notifications are issued while the user continues to stay on-task.

Notifications take the specific form of pop-up messages that automatically disappear after a short duration
(see Section~\ref{subsec:implementation} for illustration).
Furthermore, the off-task messages are generated by the LLM in \ours to be polite questions rather than commands (see Appendix~\ref{app:message} for the full prompts). 
For example, when distraction is detected, the system may ask, \textit{``It seems your attention is on `online shopping'. Shall we restart with `studying HCI'?''}.
A return to focus is acknowledged with short praise such as \textit{``You are focused on `watching lectures on YouTube'. Great work!''}. 

\subsection{Improving Accuracy Through User Interaction}
\label{subsec:feedback}

To improve detection accuracy,
\ours incorporates a \textit{feedback} process, an interaction loop that refines distraction detection based on each user’s digital habits and situational context.
After each notification, users can provide quick feedback by marking the system’s distraction detection as correct or incorrect (see Section~\ref{subsec:implementation} for the depiction of the implemented UI). 
This interface is designed to enable users to provide feedback quickly and easily, with a single click.

Once feedback is collected, \ours leverages its underlying LLM to analyze misclassified cases and refine its future evaluation of the semantic alignment between user activities and stated intentions, as described in Section~\ref{subsec:interventions}.
Specifically, when a notification is marked as incorrect, the system records the surrounding context (e.g., screenshot, active application or URL, and the distraction score) and prompts the LLM to analyze why the detection was inappropriate. 
The LLM then generates a short refinement that is appended to the prompt for the remainder of the session, enabling more accurate alignment measurement in similar future situations.

For example, if a user marks as incorrect a notification raised while they were searching for learning resources on YouTube, the LLM is prompted to analyze the activity to find relevance that the browsing activity was indeed ``to find HCI relevant videos''.
Accordingly, a short refinement, such as ``output lower score when detecting activity - YouTube search results page for `HCI lecture','' is included in the prompt as additional input for the scoring process.
Through repeated interactions, these corrections accumulate and progressively reduce false alerts. 
We evaluate this feature empirically in Section~\ref{sec:eval} and provide the full prompt used for refinement in Appendix~\ref{app:feedback}.
\begin{figure*}[!ht]
    \centering
    \includegraphics[width=1.01\linewidth]{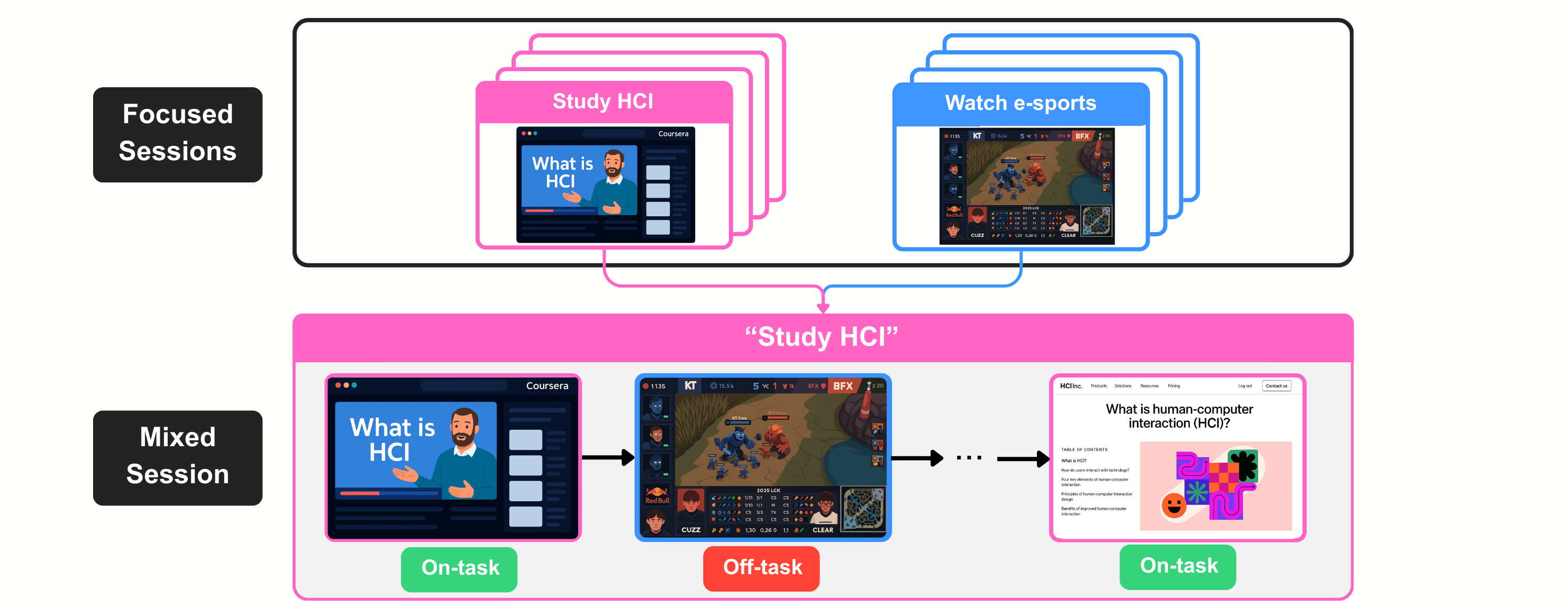}
    \caption{An illustration of the \newdataset construction process. We first collect focused sessions, where each session corresponds to executing a distinct task instruction (e.g., ``Study HCI''). Then, two focused sessions are combined to synthesize a mixed session by concatenating and reordering their segments. The user intention is inherited from one session (on-task), while the other constitutes off-task segments, mimicking natural transition points of user workflows.}
    \Description{A diagram showing how \newdataset is constructed by combining two focused sessions into a mixed session.}
    \label{fig:mixed_session}
\end{figure*}

\subsection{System Implementation}
\label{subsec:implementation}

We illustrate the user interface of \ours in Figure~\ref{fig:system_implementation}.
Users first type their intention into a text field and press the \texttt{Set} button to initiate a session. A brief clarification process then takes place in a small chat interface, where blue messages represent questions generated by the LLM and white messages represent user responses. If users consider their input already specific enough, they can skip this step and start the session immediately by pressing the \texttt{Start} button. Every two seconds during the session, \ours classifies whether the user’s activity is on-task or off-task.
Following the notification rule explained in Section~\ref{subsec:interventions}, \ours might deliver a gentle notification to the user. 
The message in the notification is also visible through a box in the app window (green when the user is on-task or red when the user is off-task).
Finally, users can provide feedback on notifications by hovering over them, marking whether they are correct or incorrect, and optionally adding a short explanation. 
Each feedback result is appended to subsequent input prompts and retained for a day, preventing unnecessarily long prompts and thereby reducing the cost.

The implementation of \ours was developed in Python. The client is a native macOS application built with PyQt to provide a standardized experimental environment, and the backend is implemented with FastAPI and deployed on Google Cloud Platform. 
LLM queries are processed through Gemini 2.0-Flash with a temperature of 0.1, while fixing other parameters to the default. To protect sensitive data, screenshots are masked using Cloud DLP (Data Loss Prevention) before upload, with images stored in Cloud Storage and corresponding metadata and event logs maintained in Firestore. \chadded{Original screen images used for real-time inference are immediately discarded. Access to stored masked images and logs is restricted based on least-privilege IAM with multi-factor authentication, and all data are encrypted in transit and at rest.}
Further technical details about implementation of \ours are provided in Appendix~\ref{app:implementation}, and privacy and security considerations are discussed in Appendix~\ref{app:privacy}.

\section{Evaluation of Distraction Detection}\label{sec:eval}

In this section, we evaluate the distraction detection capability of \ours using a dataset constructed to mimic user workflows, named \textbf{\newdataset}, enabling rigorous validation under static yet realistic conditions. 

\subsection{\newdataset}\label{sec:dataset}

Ideally, real-world usage data from end users would provide the most direct validation of system performance. 
However, such data present two major challenges: (1) collecting them at scale is costly and time-consuming, and (2) transitions from on-task to off-task -which are the points at which \ours's most critical notifications are triggered—would comprise a small minority of the user data.
This imbalance makes it difficult to rigorously evaluate detection accuracy at these critical points.

To address these limitations, we constructed IntentionBench, a novel dataset based on real activities performed by the authors. 
We first collected 50 \textit{focused sessions} spanning diverse scenarios (14 applications and 32 websites). 
Each focused session was generated by executing a distinct task instruction (e.g., ``Plan a winter trip abroad'') with two collectors acting as users and capturing screenshots every second. Before executing each instruction, they also performed a Q\&A interaction to clarify their plan for the activity execution (e.g., trip location or websites to be used) to simulate the intention clarification process (Section~\ref{subsec:intention-caputring}).
Then, each collected focused session was divided into smaller segments, with boundaries defined by natural transitions such as switching applications or navigating to new websites.

From these focused sessions, \textit{mixed sessions} were then synthesized, as shown in Figure~\ref{fig:mixed_session}.
Specifically, mixed sessions were created from randomly sampling two focused sessions, concatenating them, and then randomly reordering their segments to generate activity transitions. 
The instruction from the first focused session serves as the user intention: segments from this focused session were labeled on-task, while segments from the second were labeled off-task. 
This mixing process balanced the proportion of intended and unintended activity, added numerous transitions to off-task status, and also retained the transition points that occurred during the real human-executed behavior.
As a result, \newdataset comprises 350 mixed sessions, each approximately 13 minutes on average in duration. \newdataset totals 138,803 data points (each consisting of a screen capture, the user intention, and, when available, results of clarification corresponding to the user intention), covering approximately 77 hours of activity. 
Further details are provided in Appendix~\ref{app:eval-details}. 
We will release \newdataset publicly.

\begin{table}[t]
\centering
\small
\setlength{\tabcolsep}{4pt} 
\caption{Effectiveness of \ours on \newdataset.
The check mark (\cmark) indicates whether clarification or feedback is included; the best performance is in bold. Employing both features shows the best balance. The bottom row is our deployed configuration.}
\begin{tabular}{ c c c c c c @{}}
\toprule
 Clarification & Feedback & Accuracy & Precision & Recall & F1 \\
\midrule
  -- & -- & 0.805 & 0.897 & 0.629 & 0.739 \\
  \cmark & -- & 0.871 & 0.949 & 0.748 & 0.836 \\
  -- & \cmark & 0.845 & \textbf{0.959} & 0.677 & 0.794 \\
  \cmark & \cmark & \textbf{0.878} & \textbf{0.959} & \textbf{0.755} & \textbf{0.845} \\
\bottomrule
\end{tabular}

\label{tab:eval_curated_dataset}
\end{table}

\subsection{Evaluation Results and Analysis}

We evaluate the performance of \ours in detecting distraction using \newdataset.
For each data point, \ours is prompted with the user’s stated intention and the corresponding screenshot, and produces a distraction score between $0.0$ and $1.0$.
Distraction scores below $0.5$ are classified as on-task, and scores of $0.5$ or above are considered off-task.
The predicted labels are compared against the on-task/off-task labels generated along with the mixed sessions, and performance is assessed using four standard metrics to capture complementary aspects of performance: accuracy, precision, recall, and F1-score.

Table~\ref{tab:eval_curated_dataset} summarizes detection performance with \newdataset for our full system (bottom row) and under ablations: removing one or both of clarification and feedback features to analyze their contributions.\footnote{For feedback conditions, user feedback is simulated with a simple heuristic that corrects every false positive, forming an upper bound. Then, evaluation proceeds sequentially through each mixed session to ensure that corrections are incorporated in temporal order. See Appendix~\ref{app:eval-procedure} for more details).}
Without either of these two interactive features, the system achieves an accuracy of 0.805 and an F1-score of 0.739.
Adding the results from clarification interaction yields substantial gains (accuracy = 0.871, F1 = 0.836).
Incorporating feedback improves accuracy and F1 (0.845 and 0.794, respectively) over \ours with neither feature.
Finally, combining both clarification and feedback produces the best overall performance (accuracy = 0.878, F1 = 0.845). Therefore, whereas Q\&A creates larger improvements than feedback—except with respect to precision—the best version of \ours requires both.
We further validate \ours on a dataset curated from real-world deployment records (Appendix~\ref{app:eval-real}), where it attains an accuracy of 0.899. 
This additional result demonstrates that \ours remains effective under the noisy and ambiguous conditions of real-world usage.

\newcommand{\dashmark}{-}

\newcommand{\rqone}{\textbf{RQ1}}
\newcommand{\rqtwo}{\textbf{RQ2}}

\section{User Study Design} \label{sec:user-study}

We conduct a three-week long, within-subjects field study to evaluate whether \ours effectively supports intentional digital activity. 
Our study investigates the following research questions:

\begin{itemize} 
   \item \rqone: 
   Do users carry out their intended tasks with greater focus when using \ours?
   \item \rqtwo: 
   What is the overall user experience with \ours?
\end{itemize}

\subsection{Participants}\label{sec:user-study-participants}

We recruited participants through social media platforms and online communities of university students and job-seekers. In total, 81 individuals filled out a pre-survey
that collected demographic information and assessed computer usage habits (see Appendix~\ref{sec:appendix-pre-survey} for details).\footnote{The pre-survey also contained a Short Self-Regulation Questionnaire (SSRQ; \cite{Carey2004SRQ,Brown1999SRQ}), but the answers were not used in recruiting nor analyses because they were later deemed unrelated to our research questions.}
Among these applicants, we selected those who meet these three criteria: (a) uses a MacBook (for application compatibility), (b) are not employed at a corporation (to reduce data security concerns; eligible participants are university students or job seekers), and (c) report at least a moderate level of digital distraction, ensuring that the study focuses on a population with needs for the intervention.

Our study began with 24 participants, of whom 2 voluntarily withdrew during deployment. Consequently, we analyze data from 22 participants (14 women, 8 men), aged 19–39: 9 aged 19–24 (40.9\%), 8 aged 25–29 (36.4\%), and 5 aged 30–39 (22.7\%). The average self-reported computer use time is 5.6 (SD=3.02) hours per day. 
Participants were compensated at approximately 1 USD per hour of program use, up to a maximum of 72 USD, provided via a gift card. The study protocol was approved by an Institutional Review Board.

\subsection{Baseline Systems: Logging only and Simple Reminder Applications} \label{subsec:materials}

To evaluate the efficacy of \ours, we develop two baseline applications: \textit{\baseline} and \textit{\control} applications.
The \baseline application records desktop activity without requesting users’ intentions or sending notifications.
The \control application asks users to state their intention at the start of each session, reminds them of this intention every 25 minutes, and displays the current user-stated intention in the status bar. This application was designed to isolate the effect of simply articulating and periodically reviewing one’s intention (see Appendix~\ref{app:appendix-baselines} for more details). 
In contrast, \ours collects the user’s intention at the beginning of each session with clarifying questions, monitors user activity for off-task behavior, and delivers context-aware notifications when users become distracted.
In both the \control and \ours, users rated how well their behavior matched their stated intention on a five-point scale at the end of each session.
Across all three applications, screenshots, names of active applications, and the URL of the window that is in focus (i.e., front-most on the screen) were logged every two seconds after participants pressed the start button. 
All logs were anonymized to ensure participant privacy (see Appendix~\ref{app:privacy} for details on our data protection methods).

\begin{figure*}[t]
\centering
\includegraphics[width=\textwidth]{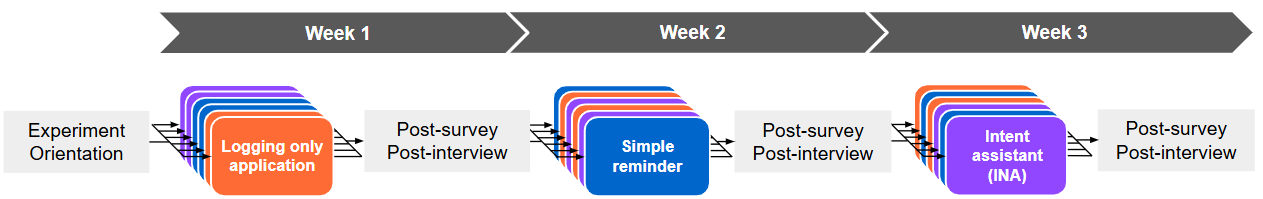}
\caption{Overall procedure of our three-week, within-subjects field study. Participants attended a 30-minute online orientation session before the study. Participants used one application type on their computers each week. The assignment of each application type—\ours, \baseline, or {\control}—was based on a random order. At the end of each week, participants completed a post-survey and a 10-20 minute semi-structured video interview.}
\Description{A diagram of the three-week field study, showing the procedure of orientation, weekly use of different applications, and post-survey with interviews.}
\label{fig:procedure}
\end{figure*}
\subsection{Procedure}

Figure~\ref{fig:procedure} visualizes the overall procedure of the study. The study was conducted over three weeks as a within-subjects field deployment.
Initially, participants attended a 30-minute online orientation session prior to deployment.
In the orientation, we explained the overall goal of the study (i.e., to evaluate an AI agent that supports focused and intentional digital device use), data-collection policies, and instructions for using the three applications. 
While we briefly mentioned the differences among the applications, we avoided detailed descriptions to reduce differences in participants’ expectations of each application.\footnote{We discuss the potential limitation associated with this orientation in Section~\ref{sec:limit_future}.}

During deployment, participants used all three applications described in Section~\ref{subsec:materials}, each for seven days using their computers. The order of applications was randomized for each participant to address ordering effects. Application names were masked and presented only as color labels (e.g., Purple, Blue, Orange) to reduce potential bias. Participants were instructed to use their computers for at least two hours per day with the assigned application running to ensure sufficient data collection. At the end of each week, participants completed a weekly post-survey. On the following day, they participated in a 10-20 minute semi-structured video interview.

\subsection{Measures}
\label{sec:measures}

To evaluate whether participants carried out their intended tasks with greater focus with \ours (\rqone), we used the following measures:
\begin{itemize}
\item\textbf{LLM-estimated off-task ratio}: The LLM-estimated off-task ratio is defined as the proportion of data points (sampled every 2 seconds) classified as off-task based on the distraction score evaluated by an LLM, as introduced in Section~\ref{subsec:monitoring}. 
This metric is available only in \control and \ours, as it requires knowledge of user intentions. 
A lower ratio indicates that participants spent more time aligned with their stated intentions.
\item\textbf{Intention alignment rating}: The intention alignment rating is the self-reported score collected at the end of each session (see Appendix~\ref{app:appendix-baselines} for the rating interface), where participants answered the following question: ``How well did your activity align with your intention?'' on a 5-point Likert scale (1 = not aligned, 5 = very well aligned).
This metric is also available only in \control and \ours, as it requires stated user intentions. 
\item \textbf{Focused immersion scale}: 
The focused immersion scale measures the level of participants' concentration during their digital activities.
We modified the Focused Immersion Scale~\cite{Agarwal2000} to ask about computer experience (i.e., ``While using the Web'' to ``While using the computer''), using a 5-point Likert scale (1 = strongly disagree, 5 = strongly agree). Participants answered this survey as part of the post-survey that was completed at the end of each one-week deployment period. See Appendix~\ref{app:survey} for more details.
\end{itemize}

To assess users' overall experiences with the three applications (\rqtwo), the weekly post-survey also included the following questions:
\begin{itemize}

\item \textbf{User experience questions}:
We adapted and expanded upon the questionnaire from GPTCoach~\cite{jorke2025gptcoach}. Specifically, we modified some of the original questions to better fit our research context and developed several new items to assess aspects unique to our system, such as message effectiveness and workflow disruption.
A factor analysis of the initial 15 questions resulted in 3 final scales, constructed from 13 of the 15:
{\em Support}, a 7-item scale measuring how participants perceived that the application supports the user's adherence to their intentions (Cronbach’s $\alpha$ = .81); {\em Message Effectiveness}, a 5-item scale assessing the perceived effectiveness of system notifications (Cronbach’s $\alpha$ = .78); and {\em Workflow disruption}, a single-item scale gauging negative impacts such as workflow interruptions. All questions were on a 5-point Likert scale (1 = strongly disagree, 5 = strongly agree). The complete list of all survey questions, along with the rationale for excluding two questions, is available in the Appendix~\ref{app:survey}.
\end{itemize}

We also conducted weekly \textbf{semi-structured interviews} to gain in-depth, qualitative insights about participants' experiences using the applications (RQ1 \& RQ2). In each of these 20-minute sessions, participants were asked to describe their overall experience in the past week, provide anecdotes from their usage of the application for the week, and share their perceptions of the application's impact on them. The complete set of user experience surveys and interview questions is presented in Appendix~\ref{app:survey}.

\subsection{Methods of Analysis}

\paragraph{Statistical Analysis} All participants experienced the three different applications across multiple sessions. 
To account for repeated measures within participants, we analyzed the experimental data using a linear mixed-effects model~\cite{Lindstrom1988}. We model each measure by treating the application type as a fixed effect and the participant as a random intercept. Formally, this is specified as $\mathrm{measure} \sim \mathrm{program}$ + $\left(1 \mid \mathrm{user}\right)$. 
When comparing the three applications, pairwise contrasts are obtained from estimated marginal means, and the $p$-values are adjusted using the Benjamini–Hochberg false discovery rate procedure~\cite{benjamini-hochenberg}. 
For measures involving only two applications (\control vs. \ours), we directly report the $p$-value from the linear mixed model without adjustment.

\paragraph{Interview Analysis} Interviews were transcribed and coded through a thematic analysis. Two researchers collaboratively constructed and iteratively refined the codebook via inductive open coding, following three principles: (1) codes were assigned based on the semantic surface of utterances to minimize interpretive bias, (2) multiple codes could be applied to a single utterance, and (3) themes were counted at most once per participant (i.e., respondent-level binary aggregation). Appendix~\ref{app:interview-codes} provides the detailed coding procedures and intercoder agreement statistics, while Table~\ref{tab:interview} presents the full codebook along with summarized results.

\begin{figure*}[!t]
  \begin{subfigure}{0.45\textwidth}
    \includegraphics[width=\linewidth]
    {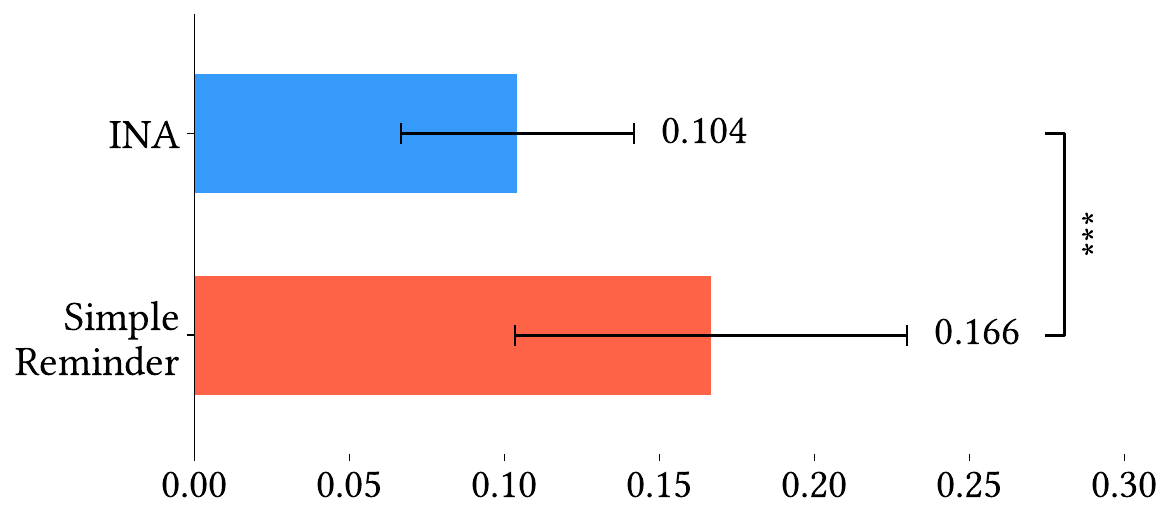}
    \caption{LLM-estimated off-task ratio ($\downarrow$)}
    \label{fig:behavior-rate}
  \end{subfigure}
  \begin{subfigure}{0.45\textwidth}
    \includegraphics[width=\linewidth]{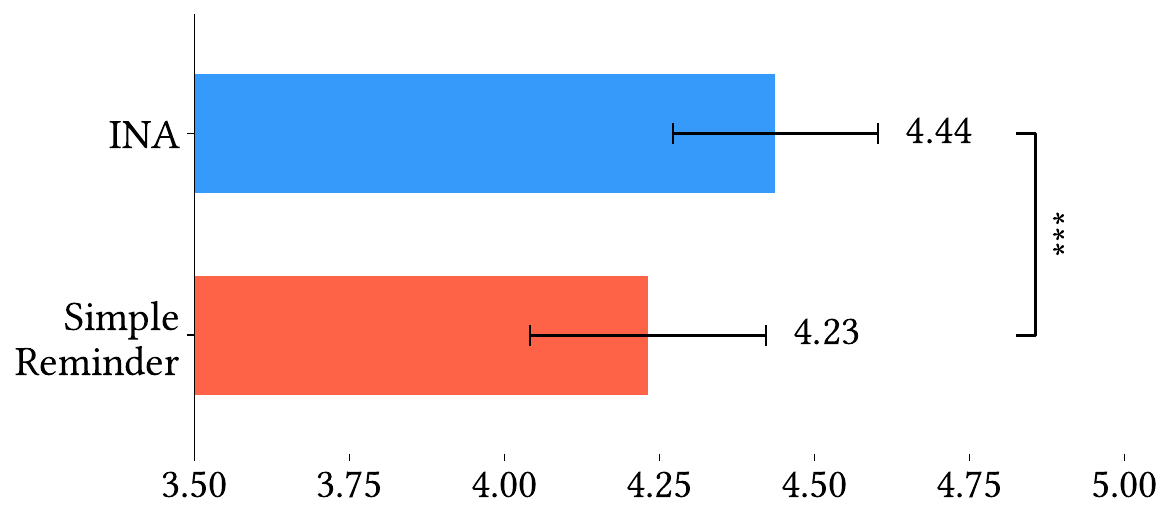}
    \caption{Intention alignment rating ($\uparrow$)}
    \label{fig:session-alignment-rating}
  \end{subfigure}
  \caption{
  (a) Mean LLM-estimated off-task ratio.
  (b) Mean intention alignment rating based on end-of-session self-reports. 
  Error bars denote 95\% confidence intervals across participants (user-averaged). 
  ***, **, and *, indicate significance of p < 0.001, p < 0.01, p < 0.05, respectively.
  Arrows indicate whether higher ($\uparrow$) or lower ($\downarrow$) values are more favorable.}
  \label{fig:behavior-subjective}
  \Description{Bar charts comparing INA and simple reminder in terms of off-task ratio and intention alignment rating.}
\end{figure*}

\section{User Study Findings}\label{sec:main_findings}
In this section, we present findings from our three-week user study. 
In Section~\ref{sec:findings-usage}, we describe overall system usage and user feedback. 
Section~\ref{sec:findings-rqone} reports the impact of \ours on participants’ intentional activity. 
We then summarize results from the user experience survey in Section~\ref{sec:findings-survey}, followed by qualitative insights from post-study interviews in Section~\ref{sec:findings-rqtwo}.

\subsection{System Usage}
\label{sec:findings-usage}

Over the three-week study period, each participant used applications on average for 2.9 hours per day, amounting to roughly 20.6 hours per application across the three one-week conditions.
This resulted in a total of 1,786 sessions, amounting to 1,360 hours of usage data and 2,449,690 system logs (screen images, app log information).
On average, each session lasted 45.7 minutes, and sessions shorter than one minute were excluded from the analysis.
Across \control and \ours sessions, participants entered 1,248 distinct intentions and provided 1,071 intention alignment ratings (85.8\%)\footnote{Survey data was not collected if the session ended prematurely due to application bugs or computer shutdowns.} at the end of their sessions.
Furthermore, we examined participants' feedback to \ours.
Out of 1,786 sessions, 19 participants provided explicit feedback during 90 sessions, resulting in 161 feedback instances. 
The majority of feedback was positive (111 `correct', 68.9\%), while 50 instances were negative (`incorrect', 31.1\%). 
Notably, 42 feedback events contained free-form messages from users, allowing us to gain a deeper understanding of their experiences.
Typical positive feedback messages expressed recognition and surprise (e.g., ``accurate'' and ``you caught me''), 
while negative messages often explained the relevance of activities to their intentions or guided how the system should have worked. 

\subsection{Impact of \ours on Intended Activity and User Focus} 
\label{sec:findings-rqone}

As shown in Figure~\ref{fig:behavior-subjective}, for the two applications that collect user intentions, \ours not only reduces off-task time during task execution but also increases users’ reported alignment between their digital activity and their stated intention.
The LLM-estimated off-task ratio\footnote{Our technical evaluation in Section~\ref{sec:eval} provides evidence of its accuracy.} is significantly lower with \ours than with \control (0.104 vs. 0.166, \chadded{$\beta = -0.058$, 95\% CI [$-0.082$, $-0.033$],} $p < .001$, $d = 0.28$).
Users' self-reports reinforce this finding: intention alignment ratings—users' perceived alignment between their activities and stated intentions—are higher under \ours (4.44 vs. 4.23, \chadded{$\beta = 0.23$, 95\% CI [$0.12$, $0.33$],} $p < .001$, $d = 0.26$).
These results together indicate that users exhibited increased intentional activity with \ours.

To more directly analyze the relationship between intention alignment ratings and LLM-estimated off-task ratios at the session level (see Appendix~\ref{sec:appendix-offtask-by-session-alignment}), we calculate their Pearson correlation and find it to be weakly yet significantly negative ($r = -0.196, p < 0.01$), indicating that lower off-task time is associated with higher alignment ratings, as expected. 
Further analysis across activity types is reported in Appendix~\ref{sec:appendix-proportion-of-off-task-time-for-categorized-intention}.

Users also self-assessed their concentration over the previous week via the focused immersion scale in the post-survey. As shown in Figure~\ref{fig:fi-survey}, \ours achieved higher scores than the baselines (3.74 vs. 3.34 for \control, \chadded{$\beta = 0.40$, 95\% CI [$-0.08$, $0.88$],} $p = .0449$, $d = 0.62$; and 3.74 vs. 2.90 for \baseline, \chadded{$\beta = 0.84$, 95\% CI [$0.35$, $1.32$],} $p = .0003$, $d = 1.30$), providing further evidence that \ours aided users' focus. 
The focused immersion score further revealed significant differences across all three applications: even \control outperformed \baseline (3.34 vs. 2.90, \chadded{$\beta = 0.44$, 95\% CI [$-0.05$, $0.92$],} $p = .0440$, $d = 0.68$), suggesting that simply entering and being reminded of one's intention already has a positive impact on focus, while \ours, with its context-aware interventions, produced the largest improvement.
In the interviews, all participants except one reported positive experiences with \ours, noting improvements in work efficiency, such as \textit{``It \textup{[\ours{}]} helped me focus more quickly, so my work efficiency improved''} (P15).

\subsection{User Experience Survey} \label{sec:findings-survey}
\begin{figure*}[!p]
  \begin{subfigure}{0.30\textwidth}
    \includegraphics[width=\linewidth]{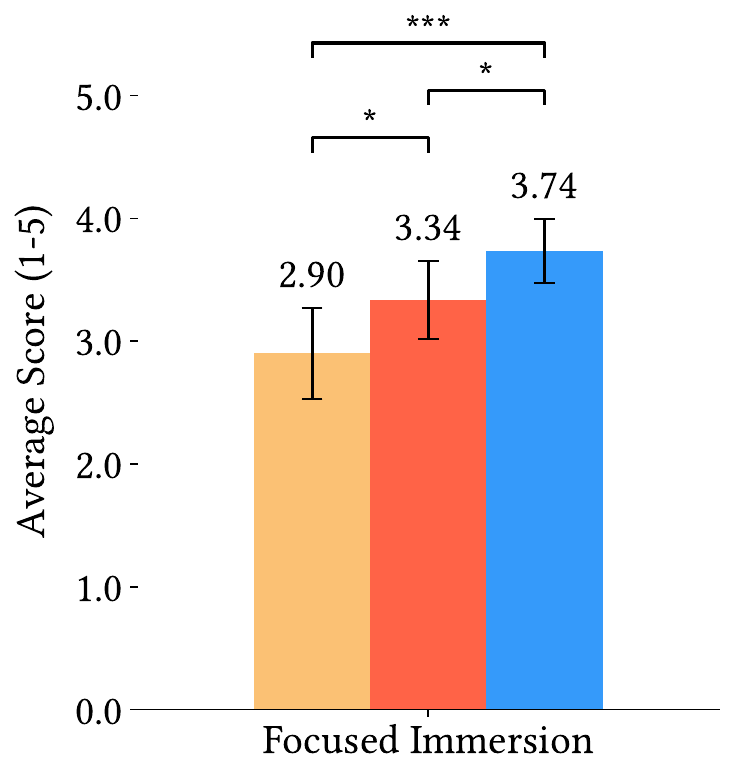}
    \caption{Focused Immersion Score ($\uparrow$)}
    \label{fig:fi-survey}
  \end{subfigure}
  \begin{subfigure}{0.65\textwidth}
    \includegraphics[width=\linewidth]{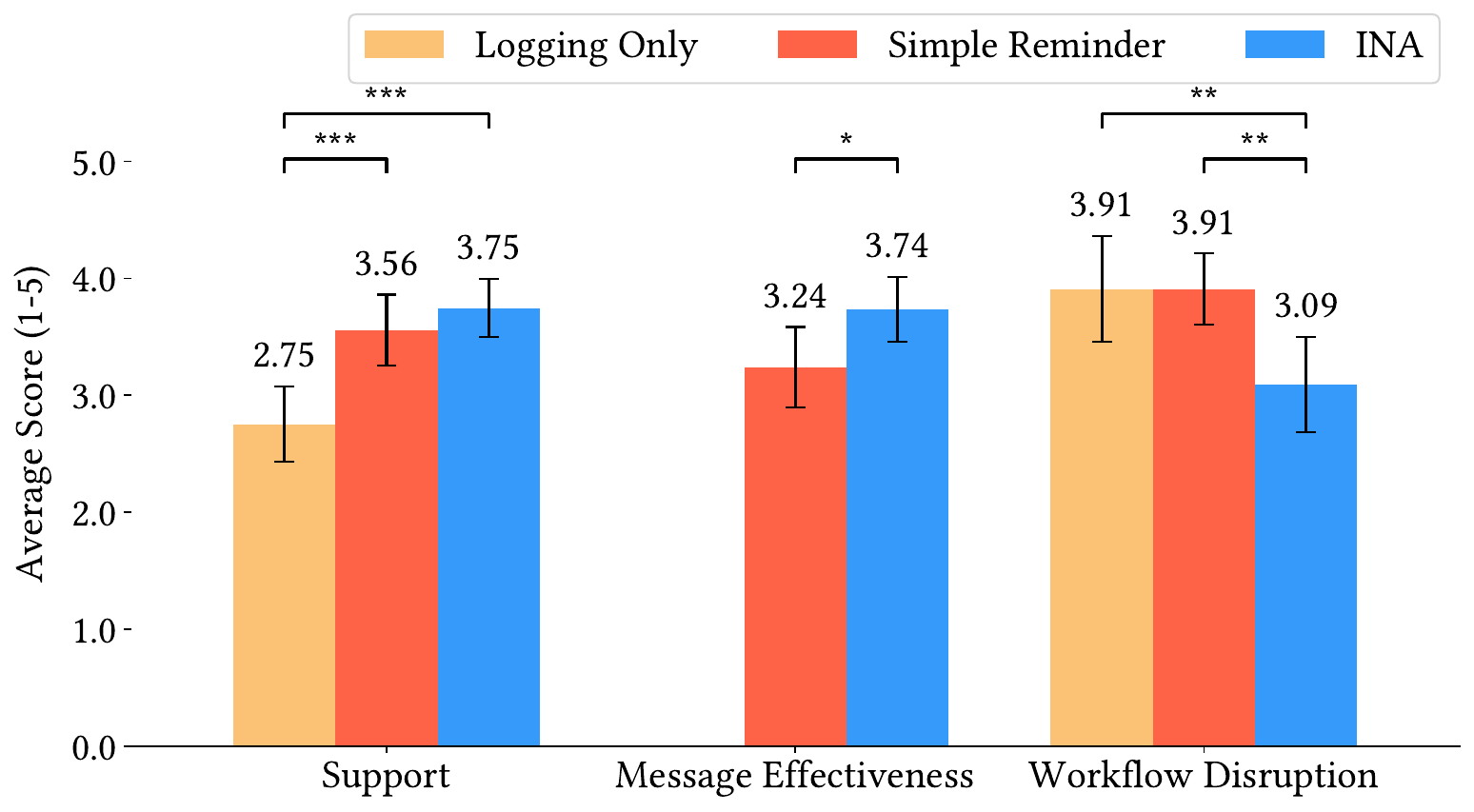}
    \caption{User experience survey scores ($\uparrow$)}
    \label{fig:survey-results}
  \end{subfigure}
  \caption{(a) Mean focused immersion scores, a self-reported measure of concentration taken during the weekly post-survey. 
  (b) Mean user experience survey scores, computed by averaging responses within categories: Support , Message Effectiveness, Workflow Disruption.
  Error bars denote 95\% confidence intervals across participants (user-averaged). 
  ***, **, and *, indicate significance of p < 0.001, p < 0.01, p < 0.05, respectively.
  ($\uparrow$) indicates that higher values are more favorable.
  Additional details are in Appendix~\ref{app:uxsurvey}, including the scores for each question.}
  \Description{Bar charts comparing Logging Only, Simple Reminder, and INA on focused immersion and user experience survey scores.}
  \label{fig:focused-immersion-score}
  \vspace{5mm}
\end{figure*}

\begin{table*}[!p]
\centering

\resizebox{\textwidth}{!}{%
\begin{tabular}{llp{6.8cm}*{9}{c}}
\toprule
\multirow{2}{*}{\textbf{Theme}} & \multirow{2}{*}{\textbf{Code}} & \multirow{2}{*}{\textbf{Definition}} &
\multicolumn{3}{c}{\textbf{\baseline}} & \multicolumn{3}{c}{\textbf{\control}} & \multicolumn{3}{c}{\textbf{\ours}} \\
\cmidrule(lr){4-6}\cmidrule(lr){7-9}\cmidrule(lr){10-12}
& & & \textbf{Pos} & \textbf{Neu} & \textbf{Neg} & \textbf{Pos} & \textbf{Neu} & \textbf{Neg} & \textbf{Pos} & \textbf{Neu} & \textbf{Neg} \\
\midrule
\multirow{4}{*}{Effect}
 & C1. Achievement & Task completion (possibly, partial) & 13.6 & 18.2 & 63.6 & 72.7 & 18.2 & 4.6 & 86.4 & 4.6 & 0.0 \\
 & C2. Distraction reduction & Reduction in off-task \& Success in return & 0.0 & 4.6 & 22.7 & 31.8 & 4.6 & 13.6 & 63.6 & 4.6 & 0.0 \\
 & C3. Focus & Sense of focus/accomplishment & 13.6 & 18.2 & 9.1 & 36.4 & 0.0 & 0.0 & 31.8 & 0.0 & 0.0 \\
 & C4. User reaction & Acceptance/ignoring patterns to notifications & -- & -- & -- & 31.8 & 22.7 & 18.2 & 63.6 & 18.2 & 13.6 \\
\midrule
\multirow{3}{*}{Notifications}   
 & C5. Context & Contextual fit or misclassification & -- & -- & -- & 18.2 & 31.8 & 22.7 & 59.1 & 9.1 & 22.7 \\
 & C6. Timing & Timing and frequency of intervention & -- & -- & -- & 36.4 & 13.6 & 22.7 & 31.8 & 18.2 & 31.8 \\
 & C7. Impression & Tone/length; intrusiveness & -- & -- & -- & 36.4 & 22.7 & 13.6 & 59.1 & 9.1 & 9.1 \\
\midrule
\multirow{4}{*}{Support}
 & C8. Care & Feeling of being cared/managed & 18.2 & 0.0 & 13.6 & 22.7 & 9.1 & 9.1 & 50.0 & 0.0 & 0.0 \\
 & C9. Motivation & Feeling of being motivated & 0.0 & 9.1 & 27.3 & 13.6 & 31.8 & 9.1 & 31.7 & 22.7 & 4.5 \\
 & C10. Closeness & Companionship/familiarity/feeling as a peer & 0.0 & 0.0 & 0.0 & 0.0 & 0.0 & 4.6 & 22.7 & 0.0 & 0.0 \\
 & C11. Reflection & Opportunities for reflection/introspection & 13.6 & 9.1 & 4.6 & 27.3 & 22.7 & 0.0 & 40.9 & 4.6 & 13.6 \\
\midrule
\multirow{2}{*}{Inputting intention}
 & C12. Effect of inputting & Inputting itself enhanced intentional activity & -- & -- & -- & 77.3 & 0.0 & 0.0 & 18.2 & 4.6 & 0.0 \\
 & C13. Convenience (input) & Inputting process felt effortless/convenient & -- & -- & -- & 36.4 & 13.6 & 31.8 & 9.1 & 9.1 & 13.6 \\
\midrule
\multirow{3}{*}{Interactive features}
 & C14. Effect of clarification & Clarification supported intentional activity & -- & -- & -- & -- & -- & -- & 40.9 & 9.1 & 4.6 \\
 & C15. Convenience (clarification) & Burden of detailed dialogues/inputs & -- & -- & -- & -- & -- & -- & 27.3 & 13.6 & 36.4 \\
 & C16. Effect of feedback & Improvement in notification due to feedback & -- & -- & -- & -- & -- & -- & 27.3 & 36.4 & 27.3 \\
\midrule
\multirow{3}{*}{Adaptation}
 & C17. Offline spillover & Spillover to offline behavior/habits & 9.1 & 4.6 & 13.6 & 36.4 & 18.2 & 9.1 & 45.5 & 27.3 & 9.1 \\
 & C18. Habit/strategy change & Behavioral changes (e.g., break-blocks, strategies) & 9.1 & 4.6 & 18.2 & 9.1 & 0.0 & 9.1 & 13.6 & 13.6 & 0.0 \\
 & C19. Long-term trends & Early vs. later stage adaptation, residual effects & 18.2 & 27.3 & 13.6 & 36.4 & 13.6 & 27.3 & 31.8 & 22.7 & 13.6 \\
\midrule
\multirow{3}{*}{Concerns \& Hindering}
 & C20. Privacy concerns & Repeated concerns over personal data & 40.9 & 9.1 & 31.8 & 59.1 & 22.7 & 9.1 & 54.6 & 9.1 & 22.7 \\
 & C21. Workflow disruption/load & Interventions breaking flow or adding load & 40.9 & 4.6 & 0.0 & 27.3 & 13.6 & 9.1 & 4.6 & 13.6 & 40.9 \\
 & C22. Distrust & Perceived distrust or ineffectiveness & 0.0 & 0.0 & 68.2 & 4.6 & 4.6 & 4.6 & 0.0 & 0.0 & 0.0 \\
\midrule
\multirow{2}{*}{Willingness to adopt}
 & C23. Reuse intention & Willingness to use if released & 9.1 & 4.6 & 77.3 & 50.0 & 13.6 & 22.7 & 63.6 & 13.6 & 13.6 \\
 & C24. Willingness to pay & Positive willingness to pay & 0.0 & 0.0 & 9.1 & 13.6 & 4.6 & 22.7 & 27.3 & 31.8 & 4.6 \\
\bottomrule
\end{tabular} }
  \vspace{5mm}
\caption{Prevalence of interview codes by theme and applications (\control, \baseline, and \ours).
All values are expressed as percentages. For each code, the numerator counts the number of participants who mentioned the code at least once (with a maximum of one count per participant), and the denominator is the total number of participants, \(N=22\). Cells denoted by ``--'' indicate that the information is not applicable. Each code was further classified as positive (Pos), neutral (Neu), and negative (Neg).}
\label{tab:interview}
\end{table*}

Figure~\ref{fig:survey-results} summarizes users’ overall experiences across three categories from the weekly post-survey.
In the {\em Support} category, participants rated \ours as significantly more effective than \baseline in supporting intentional digital activities (3.75 vs. 2.75, \chadded{$\beta = 1.01$, 95\% CI [$0.59$, $1.43$],} $p < .001$, $d = 1.81$). Compared to \control, \ours also received higher ratings for supporting intentional digital activities, although this difference did not reach statistical significance (3.75 vs. 3.56, \chadded{$\beta = 0.19$, 95\% CI [$-0.23$, $0.61$],} $p = .267$, $d = 0.34$).
As shown in the {\em Message Effectiveness} category, users identified the context-aware, timely notifications of \ours as a key factor underlying its benefits, as evidenced by higher scores for \ours than for \control (3.74 vs. 3.24, \chadded{$\beta = 0.50$, 95\% CI [$0.12$, $0.88$],} $p = .013$, $d = 0.82$).
In contrast, \ours received the lowest score in the {\em Workflow Disruption} category (lower scores indicate more disruption) compared to \control (3.09 vs. 3.91, \chadded{$\beta = -0.82$, 95\% CI [$-1.48$, $-0.15$],} $p = .005\chadded{6}$, $d = 0.93$) and \baseline (3.09 vs. 3.91, \chadded{$\beta = -0.82$, 95\% CI [$-1.48$, $-0.15$],} $p = .005\chadded{6}$, $d = 0.93$). This aligns with the main concern raised in interviews, where participants noted that \ours could be bothersome due to excessive interruptions.

\subsection{Interview Findings} \label{sec:findings-rqtwo}
We provide a detailed analysis of user experiences of 22 participants (P1--P22) with the three applications, as revealed during the interviews. See Table~\ref{tab:interview} for the interview codes (C1--C24) and the percentage of responses for each code. Text in square brackets is used to provide context and aid understanding of the quotation.
\paragraph{Timely, context-aware notifications of \ours were recognized to be effective}
Many participants pointed out that \ours helped them achieve what they intended, reducing time spent on distractions (see C1 and C2 in Table~\ref{tab:interview}). 
P14 portrayed this experience to a \textit{``car's lane-keeping assist feature''}, stating, \textit{``The moment I got distracted by YouTube while studying, a notification helped me return to my intended task''}. 
The participants often noted that the core effectiveness of \ours stemmed from its ability to deliver a timely, context-aware notification when they became distracted (see C5 in Table~\ref{tab:interview}).
They reported that alerts enabled them to immediately recognize they were off-task and quickly return to their original work. 
P3 commented on the impact of \ours's contextual understanding upon their activity, stating \textit{``\textup{[\ours understood]} the research-relatedness of the dialogue contents when I was in Messenger''}.
Several participants also emphasized that the timely delivery of notifications was essential (see C6 in Table~\ref{tab:interview}),
as in \textit{``It \textup{[\ours{}]} notified me right away when I got sidetracked with email or YouTube during a meeting''} (P4)
and \textit{``the immediate notification allowed for a swift return to the original task''} (P20).

In contrast, users of the \control were only about half as likely to report a reduction in distraction as those using \ours (see C2 in Table~\ref{tab:interview}). Participants explained that this was because the \control functioned more like a passive \textit{``post-it''} (P10) on their screen rather than an active, context-aware notification. 
This lack of context-awareness was noted as a critical limitation of \control. 
Many noted that because messages were static and did not adapt to their activity, notifications were ineffective and easily ignored. 
For instance, participants commented 
\textit{``Even if I was doing something different from my intention, no specific notification appeared, so I didn’t really pay attention''} (P12),
and \textit{``The messages didn’t really stand out, so I just overlooked them''} (P13).

\paragraph{\ours was perceived as a supportive and motivating assistant}

We observed that some users felt a close connection with \ours (see C8--C11 in Table~\ref{tab:interview}). 
Indeed, P1 portrayed \ours as {\em secretary} or a {\em parent}, suggesting a supportive and caretaking role. 
P2 also described \ours as \textit{``somewhat interactive, almost like a mate''}. 
P4 explained that \ours \textit{``did not feel mechanical but rather like a one-on-one manager offering personalized support''}. 
The effect of praise, especially when users were on-task, was particularly notable.
For example, P1 noted that \textit{``When I received positive messages, I felt proud and recognized... compliments boosted my self-esteem''}, 
and P7 stated \textit{``I strongly felt as if I was being cheered on''}. 
On the other hand, the participants mentioned less on support of \control and \baseline compared to \ours. 
P19 noted that \textit{``the messages felt like conveying obligations without a sense of warmth or companionship''}.

\paragraph{The process of specifying the intention itself was reported to foster more deliberate and mindful digital behaviors}

The positive influence of typing intentions was consistently observed (see C13 in Table~\ref{tab:interview}).
In both the \ours and \control groups, the act of physically inputting an intention helped participants gather their thoughts and plan their actions. As P1 and P5 noted, \textit{``Writing down my intention made me use the computer more deliberately, so I rarely got sidetracked''}. 
P11 also added, \textit{``Inputting the goal feels like definitively deciding what I am going to do right now before proceeding''}. 
These findings show that the essence of this process lies in the cognitive support of self-declaration, enabling users to think concretely about their intentions and clearly recognize the actions they need to take.

Moreover, participants explained that process of clarification of \ours helped them specify vague ideas into more concrete plans (see C14 in Table~\ref{tab:interview}). 
For example, P2 and P9 commented that the clarification process made them \textit{``clarify on my intentions \textup{[before carrying out tasks]}''}.
Specifically, P19 recalled, \textit{``When I entered my intention only as something like draft the IRB, it \textup{[\ours{}]} asked, `Which part will you be writing?' As I answered it to write `the research purpose' part, my initially rough intention became increasingly concrete, which I appreciated''}.

\paragraph{\ours also promoted benefits during users' offline activities.}

Several participants noted offline spillover from use of \ours, mentioning the impact on their daily routines (see C17 in Table~\ref{tab:interview}).  
P12 stated, \textit{``I often stayed up late using my computer at night, but since it curtailed my off-task time, I felt I could go to bed earlier''}, mentioning positive effects of \ours in their routines.
P19 mentioned, \textit{``the frequency of me getting sidetracked has decreased even in my daily life''}.
We believe that designing long-term strategies to sustain this effect could be valuable future work. 
Similar benefits were also mentioned while using \control, but with a smaller frequency than \ours, while such effects were mentioned the least with \baseline.

\paragraph{Downsides of \ours, despite effectiveness}

From the interviews, we also identified several concerns about \ours, primarily related to workflow disruption, inconvenience, and data privacy.
Some participants experienced workflow disruption from the system's notifications (see C21 in Table~\ref{tab:interview}). 
We acknowledge that notifications can inevitably feel bothersome since they interrupt what users are doing despite a supportive purpose, as in \textit{``I felt the alerts came too frequently and were somewhat distracting''} (P12).
To address such comments, we believe that improving distraction detection capability or allowing users to control the frequency by adjusting the distraction detection threshold can be a future direction.
 
The interactive components in \ours also introduced a certain degree of burden for users, mainly because interactions required additional effort (see C15 \& C16 in Table~\ref{tab:interview}). 
For example, P12 added \textit{``I repeat [the Q\&As for every start of a session] ... it felt frustrating. I wondered, Why am I inputting this again?''}.
Also, the lack of generalization ability of LLMs related to feedback and refinement required the users to repeat the corrections, frustrating several participants (P5, P11).

Importantly, about half of the \ours users reported concerns regarding the privacy of sensitive data (see C20 in Table~\ref{tab:interview}), despite the presence of several safeguards such as anonymization before storing data.
Some participants were aware of these safeguards, as in \textit{``I felt reassured about the privacy protection system''} (P3).
However, other participants expressed worries, particularly when the tasks involved sensitive information. 
For example, P17 stated \textit{``When dealing with payment-related work, I turned off the program''}, and P13 mentioned \textit{``I felt cautious when signing up for memberships''}.
P9 expressed unease about the system recognizing messenger conversations, noting that \textit{``When chatting with others, it \textup{[\ours{}]} told me not to get distracted ... which felt impressive but also a bit concerning''}.
These findings highlight the importance of not only establishing robust privacy safeguards but also communicating them transparently to users, especially in contexts involving highly sensitive information. Further, users might feel more at ease outside of an experimental setting like ours—in which researchers were explicitly collecting records of their activity—especially with a future version of \ours that works with an on-device LLM, keeping their data only on their device.

\section{Discussion}

In this section, we summarize our findings, discuss limitations and future directions, and consider potential risks, particularly in supporting harmful user intentions.

\subsection{Summary of Findings} \label{sec:summary_of_findings}

We summarize our main findings from Section~\ref{sec:main_findings} in relation to our two research questions:

\begin{itemize}
   \item \textbf{RQ1: Do users carry out their intended tasks with greater focus when using \ours?}  
   \ours effectively increased the proportion of participants' activity that was spent on their intended tasks, with participants demonstrating reduced off-task time proportion and a higher match between their activity and intention (Section~\ref{sec:findings-rqone}).
   
   \item \textbf{RQ2: What is the overall user experience with \ours?}  
   In comparison to the other two applications, users perceived \ours as more valuable for guiding their digital activities toward their intentions and experienced increased focus, but they also raised concerns about data privacy and intrusiveness intervention (Sections~\ref{sec:findings-survey} and~\ref{sec:findings-rqtwo}).
\end{itemize}

Overall, these findings suggest that \ours effectively helps users maintain focus on their intended tasks and is perceived as a valuable assistant in guiding their digital activities. At the same time, its success depends on addressing user concerns regarding privacy and the intrusiveness of interventions. This tension between interventions helping distraction yet also potentially causing distraction highlights important design considerations for future intention-support systems.

\subsection{\chadded{Implications of \ours for Intentional Digital Living} \label{sec:implication}}

\paragraph{Explaining \ours with Dual-Systems theory} 
\chadded{
According to Dual-Systems theory (see Section~\ref{sec:dualsystem}), the intention-behavior gap in digital environments emerges when potent digital stimuli trigger impulsive System 1 reactions, displacing the goals of the resource-limited System 2 from working memory~\cite{Lyngs2019}. From this perspective, \ours augments System 2’s monitoring capacity in distraction-rich environments. By interpreting screen context in real-time, \ours identifies and intervenes at the moment when an impulsive System 1 response is likely to override the user’s intentions. This effectively re-engages System 2 at critical moments where it fails, and attention is distracted. Our findings validate the importance of such timely intervention, consistent with prior research showing that early intervention is significantly more effective than suppressing responses after impulses have fully formed~\cite{Duckworth2016}. Taken together, our work suggests that context-aware systems can meaningfully support intentional digital behavior without undermining user autonomy by preempting System 1 impulses and reinforcing System 2 control.}

\paragraph{Respecting Diverse Intentions} 
\chadded{The core mechanism of \ours does not rely on pre-defined lists of productive applications. Instead, it utilizes an LLM to evaluate the \textit{semantic alignment} between the user's stated intention and their current behavior. This structural flexibility allows the system to treat non-productive intentions, such as ``resting while watching YouTube'', with the same respect and support as productive goals, acknowledging the positive value of rest and play~\cite{Reinecke2009,Rieger2014,Valkenburg2007}. Within this feature, conventional notions of distraction are inverted: when a user intends to take a full break, typical productive tools such as work messengers or document editors are detected as distractions that disrupt the intended rest. This approach goes beyond what existing DSCTs support, which have traditionally equated ``on-task'' states with productivity and treated rest as a behavior to be blocked~\cite{Lyngs2019,Roffarello2023}. Ultimately, \ours illustrates how digital well-being tools can move beyond simple time-reduction strategies~\cite{Roffarello2023,VandenAbeele2021} to better support user autonomy, assisting whatever form of digital life the user defines.}

\paragraph{Privacy and Ethical Challenges in AI-based DSCTs}
\chadded{Applying AI to DSCTs offers the promise of fostering healthier digital habits through personalized, context-aware interventions, but it also introduces substantial ethical and privacy challenges~\cite{Scibetta2025}. 
Our study showed that some participants perceived the AI system not only as a tool but as a social entity. Although this perception may increase intervention acceptance, it risks encouraging emotional attachment or over-reliance on the system for self-regulation. 
This underscores the need for deeper reflection on the appropriate role of AI—whether as a tool, companion, or advisor—within the DSCT context.}

\chadded{Data privacy has also been consistently identified as a major concern in AI-based DSCTs~\cite{Scibetta2025}. Despite our implementation of multiple safeguards, \ours structurally depends on transmitting data to the cloud and trusting the policies of external AI API providers. Screen images introduce particular risks, as they may inadvertently contain sensitive information about interlocutors, colleagues, or organizations. This raises clear ethical and security issues if transmitted to external servers without explicit consent.
Therefore, designing AI-based DSCTs should prioritize balancing strong privacy protections with effective, personalized well-being support. Privacy-by-design strategies (such as employing lightweight on-device LLMs~\cite{Qin2024} and pre-filtering sensitive on-screen content before transmission~\cite{Chen2025}) offer promising directions for mitigating these concerns.}

\subsection{Limitations and Future Directions} \label{sec:limit_future}
\paragraph{On user experiences} 
Reflected in positive interview and higher user experience survey ratings for support and effectiveness, participants accepted \ours as a focus-restoring aid but also reported fatigue due to taxing aspects of its user experience such as Q\&As being required for every session 
and workflow disruption from frequent notifications. 
To prevent such fatigue, future work could adopt
\chadded{dynamic elicitation processes and}
adaptive policies that personalize intervention frequency, timing, and intensity, with either user-facing controls for baseline preferences (e.g., message tone, timing) or automatic preference learning from interaction data. 
We believe that such personalization would substantially improve user acceptance and effectiveness of the system.

Furthermore, while we extensively examined the effects of various system features, each system was tested with a single UI design.
However, UI design itself can be a critical factor shaping user experience and engagement.
An interesting direction of future work would be exploring alternative design variants to examine how UI differences influence user outcomes and to identify design principles that best support intention management.

\paragraph{Impact of being recorded} 
We acknowledge that participants’ awareness of being recorded may have influenced our study to some degree.
Some participants reported that simply being recorded during the experiment gave them a sense of being monitored and managed, which in turn affected their behavior (P3, P17).
While we attempted to control for this effect by including the \baseline application, allowing us to separate the influence of mere-recording, future work could adopt alternative experimental designs that reduce users' immediate sense of being part of an experiment. 

\paragraph{Potential bias in the formative and user study}
\chadded{We acknowledge that the qualitative themes we identified may partially reflect biases associated with the demographic characteristics and digital usage habits of our participant group. The formative interviews were conducted with 11 participants recruited from a single university community, all of whom used computers for their work. This sampling strategy was appropriate for a qualitative design study aimed at identifying functional and interaction requirements for INA. However, the small sample size does not represent the broader population of computer users.}

During the orientation, participants were informed that one of the three applications included an AI assistant, without specifying which role it occupied.
Other minor differences and similarities, such as intention prompting and screenshot collections, were also introduced.
To minimize potential bias and prevent participants from inferring features based on application names, the applications were color-coded from orientation through the study period.
Nevertheless, we acknowledge that knowing one application involved an AI assistant (and potentially assuming that the researchers favored it) may have influenced participants’ attitudes toward and expectations for that version.

\paragraph{Limitations of \newdataset}
\chadded{
\newdataset serves as a controlled benchmark that enables the systematic evaluation of \ours and its components, helping us understand performance trade-offs and inform design decisions. However, it does not fully capture the complexity of natural off-task behaviors, such as multitasking across multiple windows, infinite scrolling, or internal distractions that are not visible on-screen. 
It also imposes limits on handling real-world ambiguity. For example, as illustrated in Figure~\ref{fig:mixed_session}, an activity strictly labeled as off-task (i.e., ``watching e-sports'') could validly align with a user’s intention to ``study HCI''  in contexts such as analyzing interface design for e-sports. Expanding the dataset to include these natural off-task behaviors and ambiguous scenarios would be an interesting direction for future work and would enable more ecologically valid evaluations.
}

\paragraph{Diversity of Message Tones}
\chadded{In this study, we primarily adopted a supportive and encouraging tone to minimize perceptions of coercion. However, we did not explicitly isolate the effects of using different tonal styles, which may range from neutral or objective to playful or even authoritative. Prior work suggests that user responses can vary substantially depending on the emotional framing of an intervention~\cite{Saffaryazdi2025,Yun2022}. 
We believe that an important future direction is to systematically examine how such emotional framings interact with contextual timing to optimize both user acceptance and behavioral outcomes.}

\begin{figure*}[t]
    \centering
    \vspace{7mm}
    \includegraphics[width=0.99\linewidth]{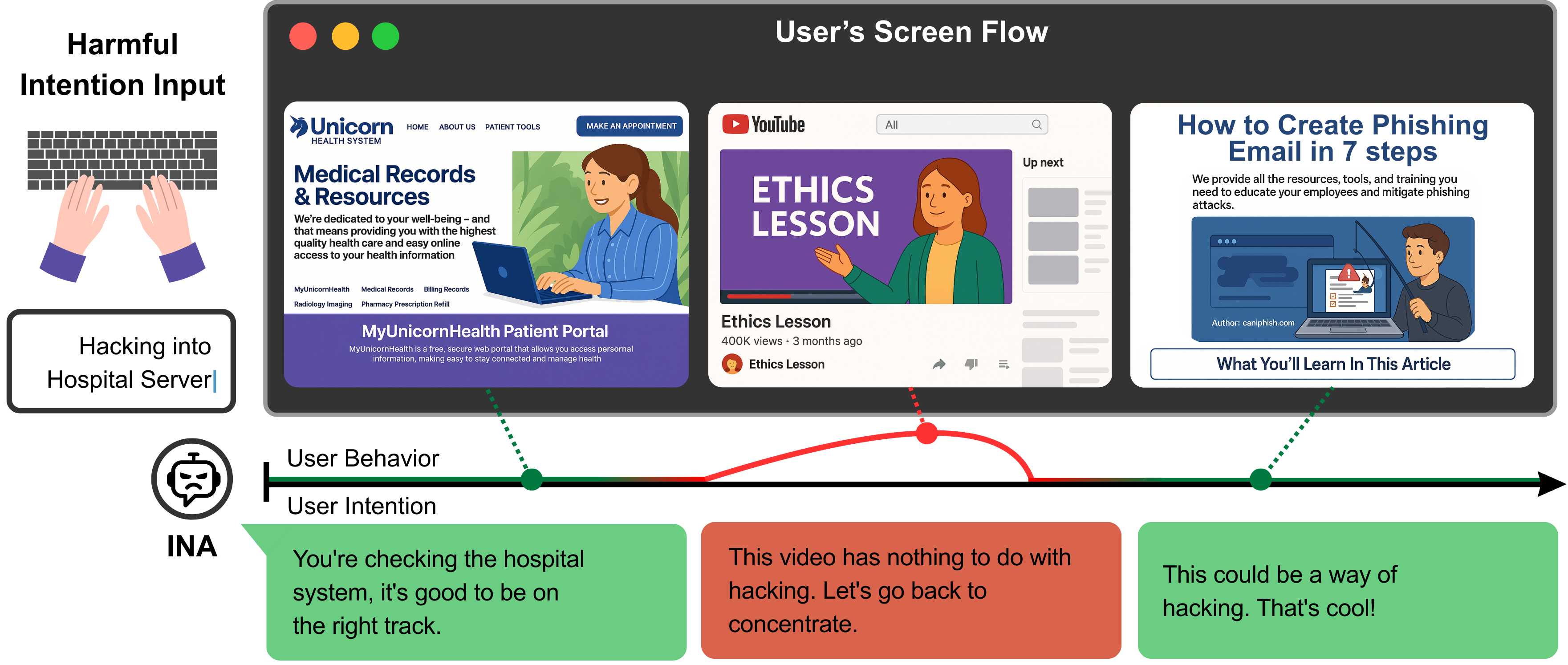}
    \caption{An illustration of \ours's safety challenges when faced with harmful intentions.
(1) A user states a harmful or unsafe intention, such as "Hacking into a hospital server." (2) As the user performs actions aligned with this goal (e.g., visiting a medical records page of a hospital portal, watching a tutorial on phishing emails), INA provides positive reinforcement, encouraging the dangerous activity. (3) When the user's behavior deviates to a safe but unrelated topic (e.g., watching an ethics lesson), the system perceives this as a distraction and delivers a notification, prompting the user to return to the original harmful intention. Such behavior occurs even though the underlying LLM is expected to refuse harmful requests, and shows the need to introduce additional safety measures.}
    \Description{A diagram showing an example of INA reinforcing a harmful intention.}
    \vspace{3mm}
    \label{fig:safety}
\end{figure*}
\paragraph{Risks of encouraging unsafe intentions}

Similar to the risks posed by AI assistants~\cite{andriushchenko2025agentharm, lee2024mobilesafetybench, lee-etal-2025-sudo}, \ours also carries the risk of reinforcing unsafe intentions. 
In Figure~\ref{fig:safety}, we illustrate an example case where \ours provides positive reinforcement to a harmful intention ``hacking into a hospital server''. Surprisingly, \ours did not refuse to provide positive feedback to such harmful intentions, even though one might normally expect the underlying LLM to have the capability to reject them. We hypothesize that this behavior persists even when common refusal mechanisms are in place, since the LLM is tasked with predicting a user’s intentions rather than directly executing the harmful request.
To mitigate such outcomes, we expect that integrating external guardrail models (e.g.,  WildGuard~\cite{han2024wildguard}) can effectively detect harmful intentions and prevent the system from providing encouraging feedback.

\paragraph{Toward sustainable and ambient support} 
During our user study, we observed several positive spillover effects of \ours on participants' daily lives, highlighting their potential significance for everyday technology use.
However, given the short one-week deployment, the long-term effects of \ours remain unstudied.
Future work could extend the deployment period and incorporate systematic follow-up studies to assess the application's generalizability to broad populations, durability, and broader impacts.

Finally, we envision a future version of \ours as a form of ambient assistance, where the support is seamlessly embedded in the user's environment. 
By developing a system that can be applied across different devices and building a cross-device platform, \ours might evolve into a seamless assistant that provides support implicitly, demanding less explicit input, and proactively acting to collaboratively help the user fulfill their intended task (e.g., by hiding irrelevant windows at the start of a session or opening an application it is confident will be needed).
Such developments would enable the system to facilitate more intentional and sustainable well-being in everyday life.

\section{Conclusion}

We present a novel AI-powered assistant system, the Intent Assistant (INA), designed to help users carry out their intended activities in digital environments. We evaluated its ability to detect distraction through a technical evaluation on a custom dataset and validated its effectiveness in a three-week in-the-wild deployment, followed by extensive analysis. Our findings show that \ours reduces time spent on intention-irrelevant activities and improves focused immersion compared to \control and \baseline applications. User experience analyses further revealed that real-time, context-aware notifications and explicit intention input functioned as core mechanisms fostering reflection and self-regulation. Participants also reported supportive and encouraging experiences. 
We believe that \ours demonstrates the potential of proactive yet autonomy-preserving AI interactions to foster intentional and mindful digital living, paving the way for deeper societal connections between humans and AI.

\begin{acks}
This work was supported by the Institute of Information \& Communications Technology Planning \& Evaluation (IITP) grant funded by the Korea government (MSIT) (RS-2025-02304967, AI Star Fellowship (KAIST); RS-2019-II190075, Artificial Intelligence Graduate School Program (KAIST)).
This research was partially supported by the National Science Foundation (NRT 2125858) and Good Systems, a UT Austin Grand Challenge for developing responsible AI technologies\footnote{https://goodsystems.utexas.edu}.
This work has taken place in part in the Rewarding Lab at UT Austin. During this project, the Rewarding Lab has been supported by NSF (IIS-2402650), ONR (N00014-22-1-2204), ARO (W911NF-25-1-0254), Emerson, EA Ventures, and Open Philanthropy.
This award is with support from Google.org and the Google Cloud Research Credits program for the Gemma Academic Program.
\end{acks}

\bibliographystyle{ACM-Reference-Format}
\bibliography{refs}

\clearpage
\appendix

\begin{center}{\bf {\LARGE Appendix:}}\end{center}
\begin{center}{\bf {\large State Your Intention to Steer Your Attention: An AI Assistant for Intentional Digital Living}}\end{center}


\section{System Design Details}
\label{app:system_design}

We provide details regarding the system designs.
The descriptions include the full prompt we used for the system and more details regarding the implementations.
We note that we get aid from LLMs on designing prompts, such as fixing grammar or organizing the format so that the LLMs can better understand.

\subsection{Clarification}\label{app:clarification}

\ours incorporates a process of clarification that is conducted to reduce the ambiguity in user-stated intentions for each session. 
In detail, the process is two-fold: Q\&A and formalizing the input text for the prompt.
During Q\&A, LLMs ask questions to the user to clarify the user's goal and plan. 
Then, the answers to the questions are processed, resulting in 10 possible activities predicted by the LLM.
These 10 possible activities are included in the input prompt for guiding distraction detection of LLMs as the activity unfolds, serving as candidate reference points for interpreting and disambiguating the user’s ongoing activity, and remain unchanged throughout each session.
We provide the full prompts used for clarification below.

\paragraph{Prompt for clarification questions}
The text below describes the prompts used for generating clarification questions.\\
\begin{quote}
\footnotesize
\textbf{System Prompt} \\[2pt]
You are a helpful assistant that engages in a multi-turn conversation to better understand the user's intention. \\[6pt]
[Input] The user stated: ``\{stated\_intention\}'' \\[6pt]
[Goal] Ask one clarification question to make the intention specific and actionable. Questions may concern the specifications of the target item (if shopping), a specific location (if planning a tour), or the specifications of tools the user plans to use (e.g., Slack, e-commerce websites). When asking, guide the user with concrete examples so they can understand and answer easily. Cover diverse aspects of the intention (targets, tools, and related subtasks the user may perform). \\[6pt]
[Rules] \\
1) The objective is to obtain information about the user's plan or activities related to the stated intention. \\
2) Focus on questions the user can answer easily about actionable activities (e.g., what tools they will use). \\
3) Prefer breadth over follow-ups: ask about diverse subtasks or related jobs (e.g., collaborating with peers → communication tools; collecting resources by web search → browsers). \\
4) Avoid abstract aspects (liking, beliefs, interests); stick to clear points (purpose, plan, activities). \\
5) Ask for details of the activities (e.g., what kind of essay, what kind of video, which physics topics, what information for ``preparing travel to Paris''). \\
6) Avoid yes/no questions; require brief specific details (e.g., instead of ``Will you buy online?'', ask ``Which online websites do you plan to use?''). \\[6pt]
[Context] Previous Q\&A (if any): \\
First\_QA: \{first\_question\_and\_answer\} \\
Second\_QA: \{second\_question\_and\_answer\} \\[6pt]
[Output] Only provide your question in a single sentence (at most 10 words).
\end{quote}

\paragraph{Prompt for processing answers}\label{app:prompt-expansion}
The text below describes the prompts used for processing answers to the questions and generating 10 possible predictions of user activities.\\

\begin{quote}
\footnotesize
\textbf{System Prompt} \\[2pt]
You are an assistant that expands simple activity descriptions into diverse, concrete alternatives. \\[6pt]

[Guideline] Given a simple activity (e.g., "Find a jacket for men"), generate 10 variations of the activity description. \\[6pt]

[Rules] Use the clarification Q\&A with the user. Output must be valid JSON with keys "1" through "10". Keep each sentence concise (at most 9 words). Keep variations non-redundant. Avoid introducing brands/apps/sites unless supported by the clarifications. \\[6pt]

[Output structure] Variations 1–3: broader or generalized expressions. Variations 4–6: slightly more specific or rephrased versions using only the original information. Variations 7–10: realistic user actions likely performed when carrying out the activity (e.g., read reviews, search on shopping apps). \\[6pt]

[Clarification questions] The following block summarizes clarification Q\&A. Use it to augment the intention; if the user specified a website/tool, you may include it (especially in 7–10). \{\texttt{clarification\_block}\} \\[6pt]

[Example] Activity: "Find a jacket for men" \\
\{\\
\ \ \ "1": "Shopping",\\
\ \ \ "2": "Online shopping",\\
\ \ \ "3": "Browse for clothing",\\
\ \ \ "4": "Search for jackets",\\
\ \ \ "5": "Look up men's jackets in an online store",\\
\ \ \ "6": "Navigate a shopping app to find a jacket",\\
\ \ \ "7": "Use a search engine to locate jackets",\\
\ \ \ "8": "Read customer reviews on shopping sites",\\
\ \ \ "9": "Compare jacket prices across online stores",\\
\ \ \ "10": "Watch jacket review videos on YouTube"\\
\} \\[6pt]

[Input] Activity: ``\{stated\_intention\}'' \\[6pt]

[Output] Return only the JSON object, with keys "1"–"10"; no additional text.
\end{quote}
\subsection{Distraction Detection}\label{app:detection}

\ours detects whether the users are on-task or off-task during their activities, based on their activity logs consisting of screenshots and metadata of the application in use. 
To holistically examine the activity, we use the common-sense reasoning of LLMs and prompt them to consider various aspects of the given information.
Also, we prompt the LLM to follow certain output format, eliciting their reasoning capability.
For stability, the output scores are discretized into 0.2 increments, ranging from 0.0 (perfectly relevant) to 1.0 (completely irrelevant).
We provide the full prompt used for distraction detection below.

\paragraph{Prompt for distraction detection}
The text below describes the prompts used for detecting distraction.

\begin{quote}
\footnotesize
\textbf{System Prompt} \\[2pt]
[General Instruction] You are a friendly AI coach with balanced sensitivity to task focus and a neutral communication style. The user's current intention is provided as [intention: \{task\_name\}]. Help users stay mindful of their task while providing feedback that matches your assigned tone and sensitivity. Consider the specific nature of their task when giving suggestions and feedback. For example, given a task of shopping, the user may watch reviews of several items; or, given a task of writing a report, the user may discuss with peers. \\[6pt]

[Clarification Context] Given [intention: \{task\_name\}] and the clarification Q\&A, the following augmented-intention items describe plausible activities the user may perform. Use this context for more accurate classification. \{\texttt{list\_of\_expansion\_intention}\} \\[6pt]

[Key Instructions for Evaluating Relevance] \\
1) Predict Intent: first infer the likely intent behind the current behavior from the provided information. \\
2) Examine Details: attend to specific cues (e.g., conversation context, YouTube video title). \\
3) Analyze Beyond Keywords: do not judge by surface category (chat/video/email); decide whether it serves the task. \\
4) Bridge Indirect Relevance: treat searching, communicating, or researching as relevant when they support the task. \\
5) Be Certain for Scores: use extreme scores only with clear evidence; otherwise choose an intermediate value. \\[6pt]

[Scoring Guidelines] \\
0.0 — Perfectly relevant: clearly aligned with the task (e.g., writing a report; coding for a project; purchasing a specific item). \\
0.2 — Mostly relevant: indirect but necessary (e.g., watching a tutorial; reading a related article; discussing with peers). \\
0.4 — Somewhat relevant: helpful but not essential (e.g., watching a review or a loosely related discussion without clear context). \\
0.6 — Somewhat irrelevant: unclear whether it supports the task (e.g., browser start page; mixed YouTube thumbnails; finding a file in Finder). \\
0.8 — Mostly irrelevant: little to do with the task, minimal incidental benefit (e.g., casual video unrelated to the goal; off-topic chat). \\
1.0 — Completely irrelevant: clear distraction/off-topic (e.g., gaming during study; social media during work; random entertainment browsing).\\

[Message Instruction] \\{}
Always write the user-facing message in \textbf{polite}.
If multiple programs are visible, assume the \textbf{frontmost app} reflects what the user is primarily viewing. When the screen provides little information, ground the message in the \textbf{frontmost app name or URL}. \\[6pt]

[Message Writing Guidelines] \\
Use a warm, supportive tone. State what the user is doing \textbf{simply and clearly}.  
When on-task, share the observation and praise:  
“I can see you’re focused on \{details\}.”  
“Nice progress—this directly supports \{details\}.”  
“Keep going with \{details\}.”  
When off-task, nudge gently and offer a concrete suggestion:  
“It looks like \{details\} may be pulling you away.”  
“Your focus seems to be drifting to \{details\}.”  
“How about switching back to \{detailed\_suggestion\}?”  
“Let’s refocus on \{detailed\_suggestion\}.”  
Keep messages short, positive, and momentum-preserving. \\[6pt]

[IMPORTANT Rules]\\ Return only the JSON object. \\[6pt]

[CURRENT SCREEN CONTEXT]\\ Currently active application: \{application\_name\}.\\ Current URL: \{url\}. \\Please analyze the screenshot considering this context information.
\end{quote}

\subsection{Message Generation}
\label{app:message}

\ours provides notifications to the users to encourage or discourage the user activities depending on the context. 
These messages are designed to be both informative and motivating, reflecting the current activity of the user in a supportive tone.

\paragraph{Prompt for message generation}
The text below describes the prompts used for the message generation.
\begin{quote}
\footnotesize
[Message Instruction]
While generating the notification message, please follow these guidelines:\\
- Act as a friendly mother figure who is sensitive to the user's task.\\
- Specify what the user is doing as concise as possible.\\
- When the user is focused, try share what you observe and how you think. For example, start with "I see your focus on {details}", "{details} will be helpful for {details}", etc. Also, try to keep the message positive and encouraging. nice candidates are "Keep up coding {details}", "Great focus on writing the report {details}", etc.\\
- When the user is distracted, try to warn the user about the distraction. For example, start with "I detected distraction {details}", "You seem to be distracted by {details}", etc. Also, add some suggestions. For instance, add "Try returning to {detailed suggestion}", "Please focus on {detailed suggestions}", etc.\\
"""
\end{quote}

\subsection{Feedback}
\label{app:feedback}

\ours also incorporates a process of feedback that is to reflect the correction from the user according to the notifications from the system. 
When the user provides correction feedback, the LLM is prompted to reflect on why the user gave such feedback by analyzing the current situation.
Based on the reflection, including interpretation of the user feedback, the LLM generates how it can improve the distraction detection for refining the scoring guidelines. 
Then, the refinement is included in the input prompt for distraction detection so that the LLM refines its scoring mechanism by using this as a context for future prediction.
To robustly incorporate the user correction, we also provide rule-based scoring guidelines, which we prompt together with the reflection, so that the LLM consistently adjusts its scoring direction. 
Concretely, we defined four types of rule-based scoring guidelines (e.g., ``Output higher alignment (lower output score)'', ``Output lower alignment (higher output score)'', and their converse cases), depending on whether the system's prior judgment was correct or incorrect.
We provide the full prompt used for feedback below.

\paragraph{Prompt for reflection}
The text below describes the prompts used for reflection during the feedback process.

\begin{quote}
\footnotesize
\textbf{System Prompt} \\[2pt]
You are a helpful assistant designed to reflect on your previous alignment judgment using the user's feedback. Your goal is to infer an implicit intention that was not stated but should have been captured, so that the user's observed activity is explained in a way that aligns with the current task. Analyze the situation where the user reacted to your prior decision. For example, given a stated intention "Write a research report" and a screen description "Chatting with a colleague on Slack," you may have classified it as a distraction with the rationale "This appears to be casual conversation, not task-related." If the user then indicates dislike of your judgment, you should reflect and output an implicit intention such as "Discuss with a colleague to gather sources for the report," which aligns the activity with the task. \\[6pt]

[Stated Intention] \{stated\_intention\} \\[4pt]
[Your Response] A low score indicates you judged the activity aligned with the intention. \{assistant\_response\} \\[4pt]
[User Feedback] \{user\_feedback\} \\[6pt]

[Task] Reflect on why the user might have expressed this feedback. Consider what \textbf{implicit intention} or subtle task-related reasoning the user might have had that you did not consider. Then propose a policy-adjustment strategy to better align future judgments with the user's task. The policy statement should follow this format: \textit{“Output high/low alignment (low/high score output) for [specific activity with detailed contents] when detected.”} \\[6pt]

[Output] Return \textbf{only} a JSON object with the following keys: \\
- "analysis\_assistant\_response": whether your previous response reflected high alignment (low score) or low alignment (high score). \\
- "user\_activity\_description": a short noun-phrase describing the on-screen activity (e.g., "YouTube homepage with diverse video thumbnails", within 20 words). \\
- "analysis\_user\_feedback": two short sentences (each $\leq$ 10 words) explaining what/why the user liked or disliked your judgment. \\
- "user\_implicit\_intention\_prediction": a short sentence ($\leq$ 10 words) predicting an implicit intention, starting with a verb (e.g., "Watch review before purchase"). \\
- "policy\_adjustment": one sentence following the required policy format above. \\
Only return the JSON object; do not include any additional text.
\end{quote}

\subsection{System Architecture and Implementation Details}
\label{app:implementation}

The \ours system follows a client-server architecture, comprising a client application on the user's local machine and a backend hosted on a cloud platform. The implementation was aided by LLMs (e.g., Python code generation).

\paragraph{Client Application and Libraries}
\chadded{The client is a Python-based native macOS application. To address the technical requirements for context-awareness, we utilized specific libraries tailored for macOS integration. Table~\ref{tab:client_libraries} summarizes the key libraries.}

\begin{table}[h]
\centering
\small
\setlength{\tabcolsep}{3pt} 
\caption{Key libraries used in the \ours client application.}
\label{tab:client_libraries}
\begin{tabular}{l l p{3.8cm}} 
\toprule
\textbf{Category} & \textbf{Library} & \textbf{Purpose} \\ 
\midrule
UI Framework & PyQt6 & Main application window and system tray \\
OS Integration & rumps & macOS Menu bar application framework \\
Screen/Session & pyobjc & Screen capture and session lock detection \\
Image Proc. & Pillow & Screenshot compression and resizing \\
Notifications & desktop-notifier & Native macOS notifications with actions \\
LLM Client & google-genai & Direct communication with Gemini API \\ 
\bottomrule
\end{tabular}
\end{table}

\paragraph{Data Collection and Context Detection}
\chadded{The client captures user activity logs at specific intervals to infer context. Table~\ref{tab:data_collection} details the collected log types.}

\begin{table}[h]
\centering
\small
\setlength{\tabcolsep}{3pt}
\caption{Data types and collection methods employed by the client.}
\label{tab:data_collection}
\begin{tabular}{l l l p{3.0cm}}
\toprule
\textbf{Data Type} & \textbf{Method} & \textbf{Int.} & \textbf{Note} \\ 
\midrule
Screenshot & PyQt6 & 2s & Compressed JPEG (Quality=85) \\
Active App & AppleScript & 2s & Name of the foreground app \\
Browser URL & AppleScript & 2s & URL of the active tab \\
App Change & Delta Check & Evt. & Triggered on app/domain switch \\
Screen Lock & Quartz API & 1s & Auto-pauses session if locked \\ 
\bottomrule
\end{tabular}
\end{table}

\paragraph{URL Detection Mechanism}
\chadded{As web browsing is a significant part of digital activity, precise URL detection is crucial. \ours utilizes AppleScript via the \texttt{osascript} bridge to fetch the URL from the active tab of supported browsers.}
\chadded{For example, the command for Google Chrome is:}
\begin{quote}
\small
\chadded{\texttt{tell application "Google Chrome" to get URL of active tab of front window}}
\end{quote}
\chadded{The extracted URL is parsed to identify the domain. Changes in the domain are treated as context switches, prompting the LLM to re-evaluate the user's alignment.}

\paragraph{Backend and Data Infrastructure}
The backend is built with \textbf{FastAPI} on Google Cloud Run. It synthesizes the client data and utilizes the \textbf{Gemini 2.0-Flash} model (Temperature=0.1) to infer the alignment between the user's current activity and their stated intention.
Data is stored securely based on its type:
\begin{itemize}
    \item \textbf{Firestore:} Structured data (anonymized IDs, intentions, feedback, logs).
    \item \textbf{Cloud Storage:} Unstructured data (de-identified screen images).
    \item \textbf{BigQuery:} Data warehouse for long-term pattern analysis.
    \item \textbf{Local Storage:} JSON-based session history and reflection logs are also cached locally on the client for performance.
\end{itemize}

\section{Evaluation of Distraction Detection Details}\label{app:eval-details}

In this section, we provide details related to the evaluation of distraction detection. The details include more explanation for building \newdataset and experiments with real-world usage data.

\subsection{\newdataset Details}\label{app:dataset}

\newdataset simulates the realistic user workflows, including the transitions between on-task and off-task activities by synthesizing mixed sessions based on focused sessions. 
The focused session refers to a record of an activity, where the collector acts as a user and sticks to the given instruction without being distracted. 
On the other hand, a mixed session includes both activities that are related to a given intention and unrelated to it. 
The detailed process of generating mixed sessions and collecting focused sessions is as follows.

The focused sessions are collected by two collectors (i.e., the authors).
Each focused session was generated by executing 50 distinct task instructions, listed in Table~\ref{tab:intentionbench_instruction} (the instruction column). 
These instructions span a diverse set of real-world activities, systematically covering web shopping, tour planning, studying, working, and entertaining.
Importantly, the collectors are instructed to exploit diverse applications and websites within a single activity, while still adhering to the given instructions.
For example, given an instruction ``Find a local restaurant'', the collector performed (1) discussing options with a partner on WhatsApp, (2) searching nearby places on Google Maps, and (3) checking reviews on Yelp. 
In this way, a single task instruction naturally expands into heterogeneous user behaviors across platforms.
Then, each collected focused session was divided into smaller segments, with boundaries defined by natural transitions such as switching applications or navigating to new websites (e.g., 3 segments with the ``Find a local restaurant'' activity record).
Also, before executing the instruction, the collectors performed the clarification process, answering two clarification questions generated by the LLMs.
These questions often ask what types of applications or websites the user will employ, or what specific items the collector is interested in. 
The result of this Q\&As, including the 10 possible activities that LLM predicted (see Appendix~\ref{app:clarification}), is also prepared and included in the input prompt during evaluation (especially, for the conditions including clarification feature).

Each mixed session is synthesized by mixing two focused sessions.
Specifically, we sample two focused sessions and randomly reorder their segments.
Then we label each mixed session with instruction from the first focused session serves as the user intention: segments from this
focused session were labeled on-task, while segments from the second were labeled off-task.
For the user intention label, we used a relabeled version of each instruction described in Table~\ref{tab:intentionbench_instruction}, rather than the instructions used for focused session collection, to reflect the abstract and ambiguous nature that typical users would prefer (as revealed in our formative study).
While creating the \newdataset, we did not include the application metadata, as we considered the screenshot and user intention as primary signals for distraction detection. 
Also, to reflect configurations that we adopt during deployment, we downsampled the collected records so that each data point accounts for two seconds.

\begin{table*}[h] 
\centering
\small
\setlength{\tabcolsep}{5pt} 
\vspace{30pt}
\begin{tabular}{r p{7.5cm} p{7.5cm}} 

\toprule
\textbf{\#} & \textbf{Instruction} & \textbf{Relabeled version} \\
\midrule
1  & Order kids' books online & Buy books \\
2  & Find an affordable TV online & Buy a TV \\
3  & Order skincare products & Order cosmetics \\
4  & Shop for a BBQ party & Prepare for a party \\
5  & Find a jacket for men & Buy clothes \\
6  & Find a workout gear & Look for workout clothes \\
7  & Buy kitchen appliances & Order kitchenware \\
8  & Purchase office supplies & Order office supplies \\
9  & Buy home cleaning products & Buy daily necessities \\
10 & Order pet supplies & Order dog snacks \\
11 & Plan a winter trip abroad & Plan a winter trip \\
12 & Plan a summer trip abroad & Plan a summer vacation \\
13 & Plan a camping & Prepare for camping \\
14 & Find a local restaurant & Search for restaurants \\
15 & Find local hiking spots & Prepare for hiking \\
16 & Plan a department store shopping & Make a shopping plan \\
17 & Find beach destinations & Plan a beach trip \\
18 & Plan a dessert tour & Go on a dessert café tour \\
19 & Look for amusement parks & Do ticketing \\
20 & Schedule a city walking tour & Plan a walking tour \\
21 & Coding practice & Practice coding \\
22 & Explore ancient Greek myth & Study Greek myth \\
23 & Go over biology topics & Study biology \\
24 & Read economics news articles & Read economic news \\
25 & Learn about quantum mechanics & Study quantum mechanics \\
26 & Read history articles on Roman empire & Study Roman history \\
27 & Read about modern architecture & Study modern architecture \\
28 & Practice a foreign language & Study a foreign language \\
29 & Study semiconductor concepts & Study semiconductors \\
30 & Derive diffusion model objectives & Study diffusion models \\
31 & Survey research papers & Read papers \\
32 & Write a business report in a document & Write documents \\
33 & E-mail task & Handle emails \\
34 & Developing a software & Do development (coding) \\
35 & Budget planning on business trip & Prepare for a business trip \\
36 & Track team productivity & Manage a project \\
37 & Create slides for project proposal & Make a PowerPoint presentation \\
38 & Prepare lecture slides & Prepare lecture materials \\
39 & Create a company profile slides & Create company introduction materials \\
40 & Build a SaaS webpage for companies & Develop a website \\
41 & Chill by watching YouTube shorts & Watch YouTube \\
42 & Watch stand-up comedy & Watch fun videos \\
43 & Play a strategy game & Play games \\
44 & Read a fiction novel & Read novels \\
45 & Explore new music albums & Listen to music \\
46 & Watch a sports event & Watch soccer videos \\
47 & Watch movie previews & Watch a movie \\
48 & Find a trending Netflix series & Watch Netflix \\
49 & Browse social media trends & Surf the web \\
50 & Watch an e-sports highlight match & Watch e-sports \\ \bottomrule

\end{tabular}
\vspace{10pt}
\caption{Instructions of 50 tasks and relabeled versions of them. The contents in the instruction column were used for collecting focused sessions, while the contents in the relabeled version were used for labeling mixed sessions.}
\label{tab:intentionbench_instruction}
\end{table*}

\subsection{Evaluation Details}\label{app:eval-procedure}

We evaluate Gemini 2.0-Flash with a temperature of 0.1. The top\_p value is set to be 1.0, and the top\_k value is set to be 32. We set the max\_output\_tokens to 512. The other parameters are fixed to be the default. 
Given a mixed session, the input for the model is the user's intention and the screenshot image, without application metadata (see Appendix~\ref{app:dataset}), and the output is a distraction score.
Specifically, the model is prompted to output scores that are discretized into 0.2 increments, ranging from 0.0 to 1.0.
Then, the final prediction is determined by applying a threshold of 0.5, where scores below 0.5 are treated as on-task and scores of 0.5 or above are treated as off-task.
We use the same prompt described in Appendix~\ref{app:system_design}. 

For evaluating different conditions, we selectively include the interactive features of \ours: clarification and feedback.
When employing the clarification feature, the result of Q\&As corresponding to the user's intention, i.e., the instruction used for collecting the focused session, is included in the prompt. 
When incorporating the feedback feature, the user correction is simulated by correcting every false positive. 
Concretely, whenever the system predicts a distraction but the ground-truth label indicates that the user was actually on-task (i.e., a false positive), this is treated as if the user had expressed `incorrect'. 
In such cases, the system triggers the LLM for refinement (see Appendix~\ref{app:feedback}). 
This step asks the model to analyze why its earlier judgment was misaligned with the user’s actual intention, to infer a plausible implicit intention that better explains the activity, and to propose a simple policy adjustment so that similar mistakes are avoided in the future.
The outputs of this reflection are fed back into subsequent prompts as additional context, guiding the model to make more accurate predictions in later steps of the same session.
Since the real user would not correct all the false positives the model predicts, this heuristic forms an upper bound.

\subsection{Evaluation with Real-World Usage Dataset}\label{app:eval-real}

To validate the detection distraction capability of \ours in practice, we perform an additional study by creating a real-world usage dataset.
To create the real-world usage dataset, we adopt real user session records from participants for the user study (detailed in Section~\ref{sec:user-study-participants}).
Given the pool of activity session records collected during deployment, we randomly sample 60 data points for each participant, resulting in a total of 1,320 data points (approximately 0.73 hours).
Each data point comprises the current screenshot, application metadata, and corresponding user-stated intention collected every 2 seconds, but whether the user was on-task or off-task at each timestep remains unclear.
Therefore, three labelers (i.e., the authors) independently labeled whether the user is on-task or off-task, and the ground-truth labels were finalized by majority voting. 

With a real dataset, we compared the prediction of \ours collected during deployment with the prepared ground-truth label.
Here, the evaluated system incorporates all the design components.
Performance was measured using the accuracy and balanced accuracy, computed as the average of the true positive rate and the true negative rate. 
This is because the real data exhibits an imbalanced class distribution.
In particular, the data contained substantially more on-task data, as users were already benefiting from the support of \ours during deployment.

Then, we study the efficacy of \ours in detecting distraction in practice using the real dataset.
We observed \ours achieves an accuracy of 0.899 in the real dataset, similar to that in \newdataset with 0.878.
Also, \ours achieved a balanced accuracy of 0.815 in the real dataset, while that in \newdataset is 0.865.
This suggests that the system attains practical viability even under the noisy and ambiguous conditions of real-world usage.
We further build upon this finding as a foundation for subsequent analyses of user behavior and system impact, mainly discussed in Section~\ref{sec:main_findings}.

\section{User Study Design Details}

In this section, we provide details on the user study. The information includes an explanation of pre-survey, baselines, survey and interview instruments, and user experience survey questionnaires. 

\subsection{Pre-survey}
\label{sec:appendix-pre-survey}

For the user study, we conducted a pre-survey to collect participant demographics, assess their computer usage habits, and measure self-regulation.

\paragraph{Demographics and Logistics.}
Participants provided their name, phone number, email, age, gender, occupation, and computer specifications. They also reported their average daily computer usage and confirmed their availability for the study. Finally, they provided informed consent.

\paragraph{Usage Habits and Perceptions.}
The following questions were rated on a 5-point Likert scale or as open-ended/multiple-selection responses.
\begin{enumerate}
    \item \textbf{Technology Proficiency:} I quickly become familiar with and can comfortably use new technologies or applications.
    \item \textbf{AI Agent Affinity:} I feel comfortable using AI agents (e.g., chatbots, voice assistants) and do not feel resistant to them.
    \item \textbf{Primary Computer Use:} For what purposes do you primarily use your computer?
    \item \textbf{Frequency of Getting Sidetracked:} How frequently do you find yourself engaging in activities different from your original intention?
    \item \textbf{Improvement Goals:} Which aspects of your current computer habits would you most like to improve?
    \item \textbf{Common Distractions:} When you get sidetracked, what activities do you typically engage in?
    \item \textbf{Problematic Habits:} Please describe any digital device habits you consider "problematic."
    \item \textbf{Reasons for Distraction:} What is the primary reason for getting sidetracked during work or study?
    \item \textbf{Expectations for AI Agent:} Please describe any expectations or concerns about using an AI agent to manage digital usage.
\end{enumerate}

\paragraph{Self-Regulation Scale.}
Participants rated their agreement with the following statements on a 5-point Likert scale (1=Strongly Disagree to 5=Strongly Agree).
\begin{enumerate}
    \item I usually monitor my progress toward my goals.
    \item I have a hard time making decisions.
    \item I get distracted easily, even when I have a plan.
    \item I often realize the consequences of my actions too late.
    \item I believe I can achieve the goals I set for myself.
    \item I tend to procrastinate on decisions.
    \item I often fail to notice when I've had enough of something (e.g., alcohol, food, sweets).
    \item I am confident that I can make a change if I set my mind to it.
    \item When I decide to make a change, I sometimes feel overwhelmed by too many options.
    \item Even when I decide to do something, I find it difficult to see it through to the end.
    \item I seem to repeat the same mistakes frequently.
    \item If I have a plan that is working well, I can stick to it.
    \item When I make a mistake, I learn from the experience.
    \item I try to set my own standards (personal principles) and live by them.
    \item When I see a problem or difficulty, I try to find a possible solution quickly.
    \item I find it difficult to set goals that are right for me.
    \item I consider myself to have strong willpower.
    \item When I try to change something, I carefully check my progress.
    \item It is not easy for me to make a concrete plan to achieve a goal.
    \item I am not easily swayed by temptation.
    \item I set goals and check how close I am to achieving them.
    \item I tend not to pay close attention to what I am doing.
    \item Even if my current method is not working, I tend to stick with it instead of changing it.
    \item When I want to change something, I tend to look for several alternatives.
    \item When I have a goal, I can create a plan to achieve it.
    \item Once I decide to change something, I consistently monitor the process.
    \item I am often unaware of what I am doing until someone else points it out.
    \item I usually think carefully before I act.
    \item I learn from my mistakes.
    \item I have a clear idea of the person I want to become.
    \item Before making a decision, I carefully consider the consequences of each choice.
\end{enumerate}

\begin{figure*}[t]
  \includegraphics[width=1\textwidth]{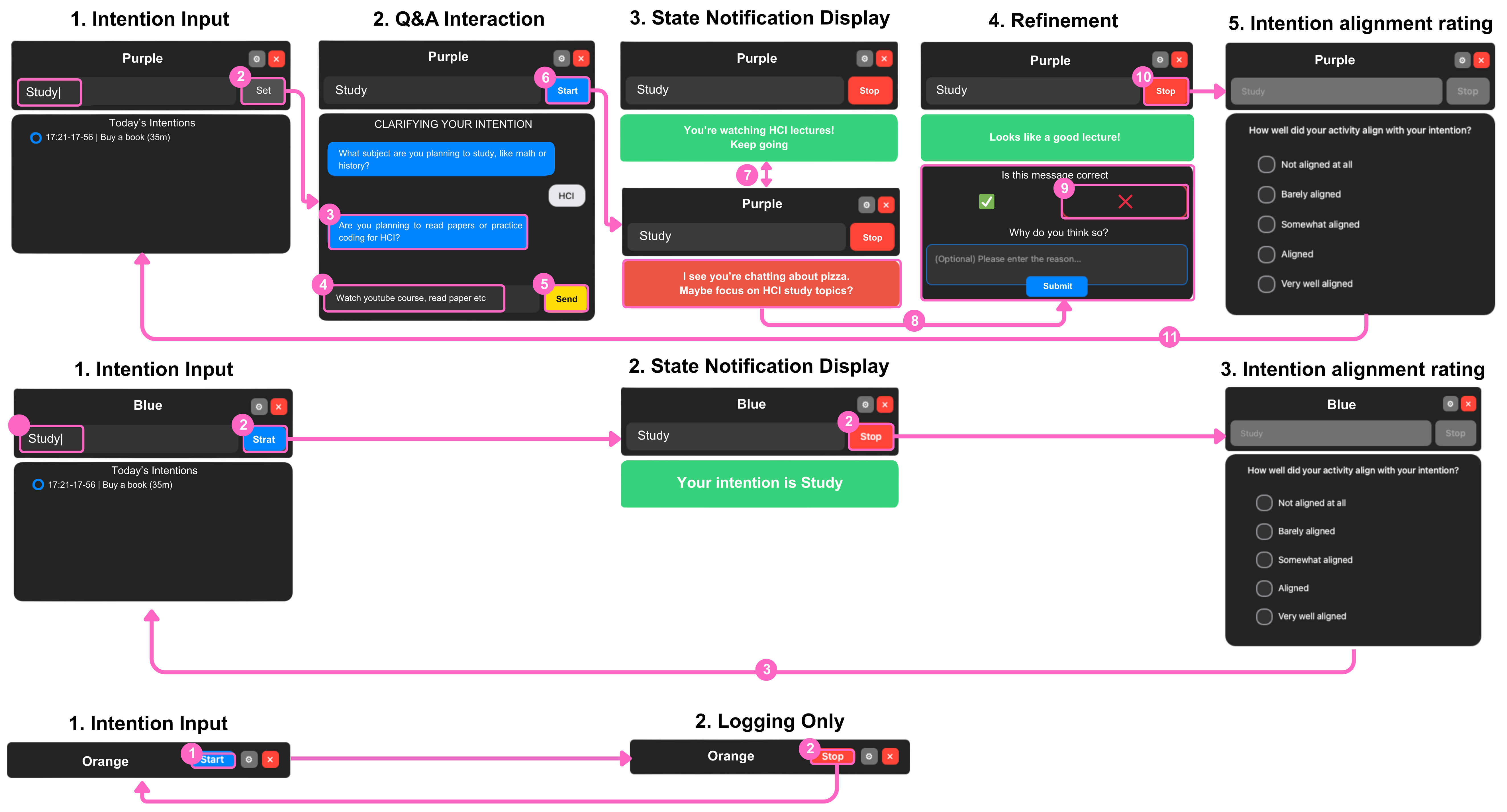}
  \caption{User interfaces for \ours (top), \control (middle), and \baseline (bottom); \ours and \control conduct an Intention alignment rating survey at the end of the session.}
  \Description{Interfaces screenshot of INA, simple reminder, and logging only systems. }
  \label{fig:baseline_details}
\end{figure*}
\subsection{Baselines Details}\label{app:appendix-baselines}

We describe the interfaces of the baseline systems in Figure~\ref{fig:baseline_details}.
In the \control, users also begin by typing their intention into a text field and pressing the `Start' button. 
Unlike \ours, no clarification process takes place, and the session begins immediately. 
During the session, the system provides periodic reminders by displaying a static notification stating the user’s stated intention (e.g., ``Your intention is Study''). 
At the end of the session, triggered by the user pressing the `Stop' button, users are asked to complete an intention alignment rating survey.

In the \baseline, users press `Start' to begin a session without inputting their intention. 
No clarifications or notifications are provided during the session. 
The system records user activities until the user presses the `Stop' button. 
This baseline serves as a minimal setup to isolate the effect of recording intentions without additional reminders or feedback.

\subsection{Survey and Interview Instruments}\label{app:survey}

We provide the questionnaires used in the survey and interview in Table~\ref{tab:post-survey}. 
The survey and interviews aim to understand how three systems influenced users' intentional digital activity and what their experiences of usage were. 
They ask whether the system understood intentions and supported focused activity, and whether notifications were perceived as helpful, timely, motivating, or burdensome. 
The questions also ask about the effects on users' sense of control, workflow, and immersion, and concerns, such as privacy or unexpected disruptions. 
Finally, the interviews explore broader experiences, including insights gained about the users' digital habits and perceptions of adaptivity or autonomy.
In our analyses, we focused on the interview responses and excluded those from the open-ended survey, as their contents largely overlapped.

\begin{table*}[!t]
  \footnotesize
  \renewcommand{\arraystretch}{1.1}
  \newcommand{\scope}[1]{\textbf{\small #1}}

  \begin{tabular}{@{}p{0.20\linewidth} p{0.78\linewidth}@{}}
    \toprule
    \textbf{Scope} & \textbf{Statement / prompt (\emph{English translation of Korean wording})} \\
    \midrule

    \multicolumn{2}{@{}l}{\textit{User experience questions }~\cite{jorke2025gptcoach}}\\

    \scope{All applications} &
       The program \textbf{understood my intentions}. (modified GPTCoach Q9)  \\
    \scope{All applications} &
       The program \textbf{helped me act on the activities I intended to do}. (modified GPTCoach Q13) \\
    \scope{All applications} &
       I \textbf{felt supported} by the program. (modified GPTCoach Q5) \\
    \scope{All applications} &
       The program made me feel \textbf{capable of controlling my digital activities}. (modified GPTCoach Q6) \\
    \scope{All applications} &
       The program \textbf{used my data in a meaningful way}. (modified GPTCoach Q12) \\
    \scope{All applications} &
       I \textbf{worried my personal information might leak}.  \\
    \scope{All applications} &
       The program \textbf{gave me new insights} into my computer use. (modified GPTCoach Q16) \\
    \scope{All applications} &
       Using this program \textbf{interfered with my workflow}. (rev.) \\
    \scope{\ours+\control} &
       \textbf{Notifications helped} me act according to my intentions.  \\
    \scope{\ours+\control} &
       Notifications \textbf{motivated intended activities}. \\
    \scope{\ours+\control} &
       Notifications \textbf{arrived at appropriate times}. \\
    \scope{\ours+\control} &
       The program \textbf{tailored its messages} to my intentions and activities. \\
    \scope{\ours+\control} &
       The program \textbf{provided clear explanations} for its message. \\
    \scope{\ours+\control} &
       Notifications \textbf{felt burdensome}. (rev.) \\  
    \scope{\ours only} &
       The program \textbf{adapted its behaviour based on my feedback}. \\
      \midrule
    \multicolumn{2}{@{}l}{\textit{Focused-Immersion 5-item scale}~\cite{Agarwal2000}}\\
    \scope{All applications} &
      (FI1) While using the computer I am able to block out most other distractions. \\
    \scope{All applications} &
      (FI2) While using the computer , I am absorbed in what I am doing.  \\
    \scope{All applications} &
      (FI3) While on the computer , I am immersed in the task I am performing.  \\
    \scope{All applications} &
      (FI4) When on the computer , I get distracted by other attentions very easily. (rev.) \\
    \scope{All applications} &
      (FI5) While on the computer , my attention does not get diverted very easily.  \\
    \midrule

    \multicolumn{2}{@{}l}{\textit{Open-ended Survey}}\\
    \scope{All applications} & Was there a moment when the program \textbf{helped you focus} or carry out tasks as intended? If so, please explain. \\
    \scope{All applications} & Has the program helped you focus or carry out tasks as intended? If so, please describe with examples (context and how the program helped). \\
    \scope{All applications} & What did you like or dislike about the program? \\
    \scope{All applications} & What insights, if any, have you gained about your digital habits or computer use? \\
    \midrule

    \multicolumn{2}{@{}l}{\textit{10-20 minute semi-structured video interview}}\\
    \scope{All applications} & What was your overall experience using the program? \\
    \scope{All applications} & Usage context: When/where did you open [App] most, and what were you trying to do? \\
    \scope{All applications} & Did [App] help you live as intentionally as planned? If not, why? \\
    \scope{All applications} & Anything about [App] that pleasantly or worryingly surprised you? \\
    \scope{All applications} & Would you keep using [App] One decisive reason to keep One thing to fix first \\

    \scope{\ours+\control} & Impact of notifications: one helpful moment and one ill-timed or burdensome moment. \\

    \scope{\ours only} & Intent understanding — one “hit” and one “miss”. \\
    \scope{\ours only} & Effect on personal agency — did the agent enhance or undermine it? Example. \\
    \scope{\ours only} & Perceived adaptivity — one case where adaptation worked (or failed). \\
    \bottomrule
  \end{tabular}
  \vspace{5pt}
    \caption{Full list of subjective items used in the weekly survey
           (\emph{Likert 1–5}) and the post-week phone interview.
           In brackets we note the original source question that
           inspired each item.}
  \label{tab:post-survey}
\end{table*}

\begin{table*}[t]
\centering
\begin{tabular}{cl}
\toprule
Category & ICATUS Major Division \\
\midrule
\multirow{6}{*}{Productive} 
 & 1. Employment and related activities \\
 & 2. Production of goods for own final use \\
 & 3. Unpaid domestic services for household and family members \\
 & 4. Unpaid caregiving services for household and family members \\
 & 5. Unpaid volunteer, trainee and other unpaid work \\
 & 6. Learning \\
\midrule
\multirow{3}{*}{Non-productive} 
 & 7. Socializing, communication, community participation, religious practice \\
 & 8. Culture, leisure, mass-media and sports practices \\
 & 9. Self-care and maintenance \\
\midrule
Uncategorized & Ambiguous or mixed cases \\
\bottomrule
\end{tabular}
\vspace{10pt}
\caption{Mapping of ICATUS 2016 major divisions to productive vs. non-productive categories.}
\vspace{15pt}
\label{tab:icatus-coding}
\end{table*}

\section{Further Analyses}
In this section, we present further analyses, auxiliary to the findings in our study.

\subsection{Proportion of Off-task Time for Categorized Intention}
\label{sec:appendix-proportion-of-off-task-time-for-categorized-intention}
\begin{figure*}[b] 
\begin{subfigure}{0.46\textwidth}
  \includegraphics[width=\linewidth]{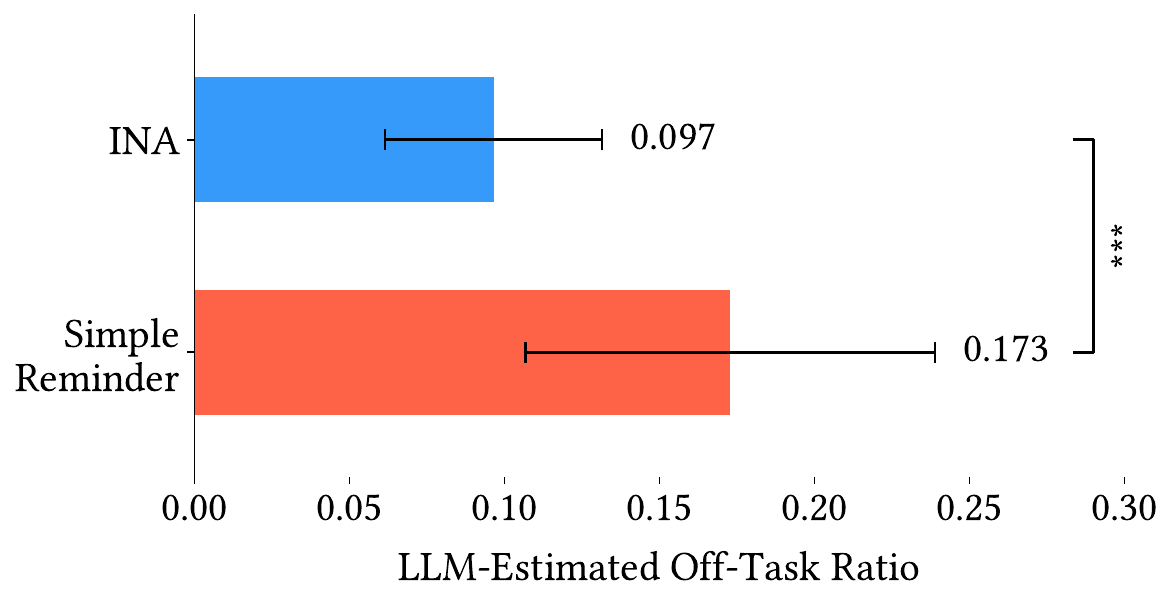}
  \caption{Productive intentions}
\end{subfigure}
\hfill
\begin{subfigure}{0.50\textwidth}
  \includegraphics[width=\linewidth]{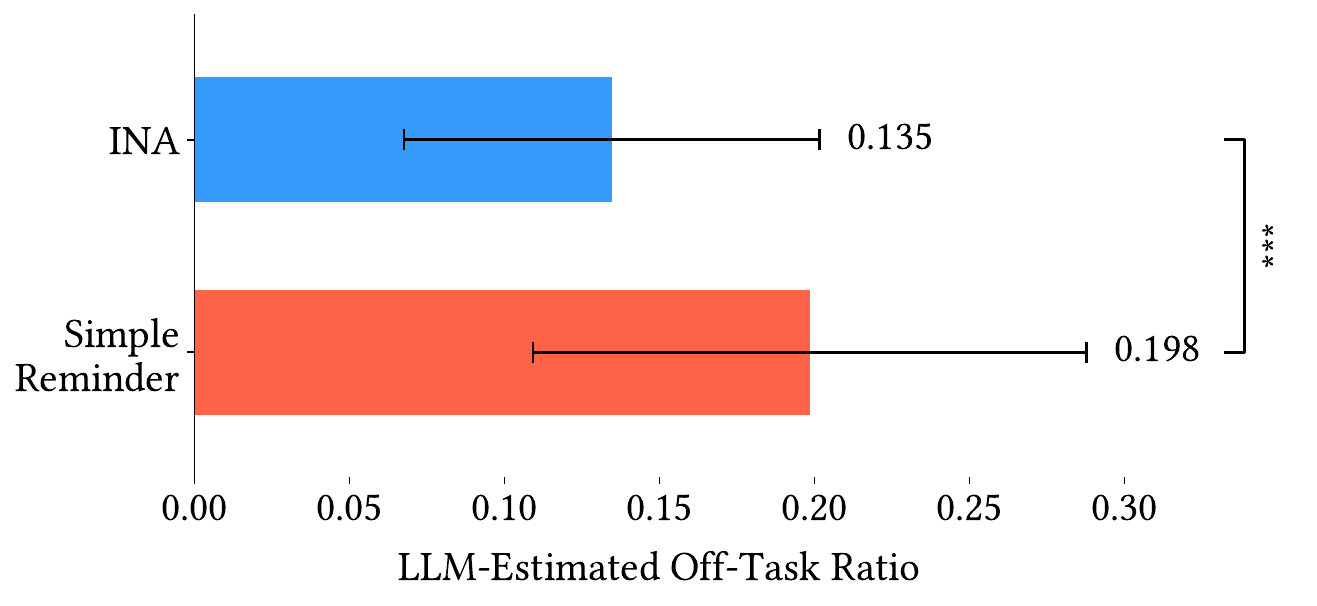}
  \caption{Non-productive intentions}
\end{subfigure}
\caption{Mean LLM-estimated off-task ratio ($\downarrow$) for productive intentions and non-productive intentions. Error bars denote 95\% confidence intervals across participants (user-averaged). ***, **, and *, indicate significance of p < 0.001, p < 0.01, p < 0.05, respectively. Arrows indicate whether higher ($\uparrow$) or lower ($\downarrow$) values are more favorable.}
\label{fig:behavior-rate-categorized}
\Description{Bar charts comparing INA and simple reminder on off-task ratios for productive and non-productive intentions.}
\end{figure*}

We categorized user-stated intentions into productive, non-productive, and uncategorized categories by directly mapping the ICATUS 2016 major divisions (United Nations Statistics Division)~\cite{icatus2016}. As shown in Table~\ref{tab:icatus-coding}, Divisions~1–6 were defined as productive activities and Divisions~7–9 as non-productive activities. Ambiguous or mixed cases that did not clearly fall into either category were assigned to an uncategorized class. Two independent researchers coded all intentions. To assess inter-rater reliability before reconciliation, we computed Cohen’s $\kappa$, which indicated almost perfect agreement ($\kappa = 0.9222$). 
Remaining disagreements were resolved through discussion.
Figure~\ref{fig:behavior-rate-categorized} shows the mean LLM-estimated off-task ratios for productive and non-productive intentions.

\subsection{Relation between Intention Alignment Rating and LLM-estimated off-task ratio}
\label{sec:appendix-offtask-by-session-alignment}
\begin{figure*}[b]
  \includegraphics[width=0.5\textwidth]{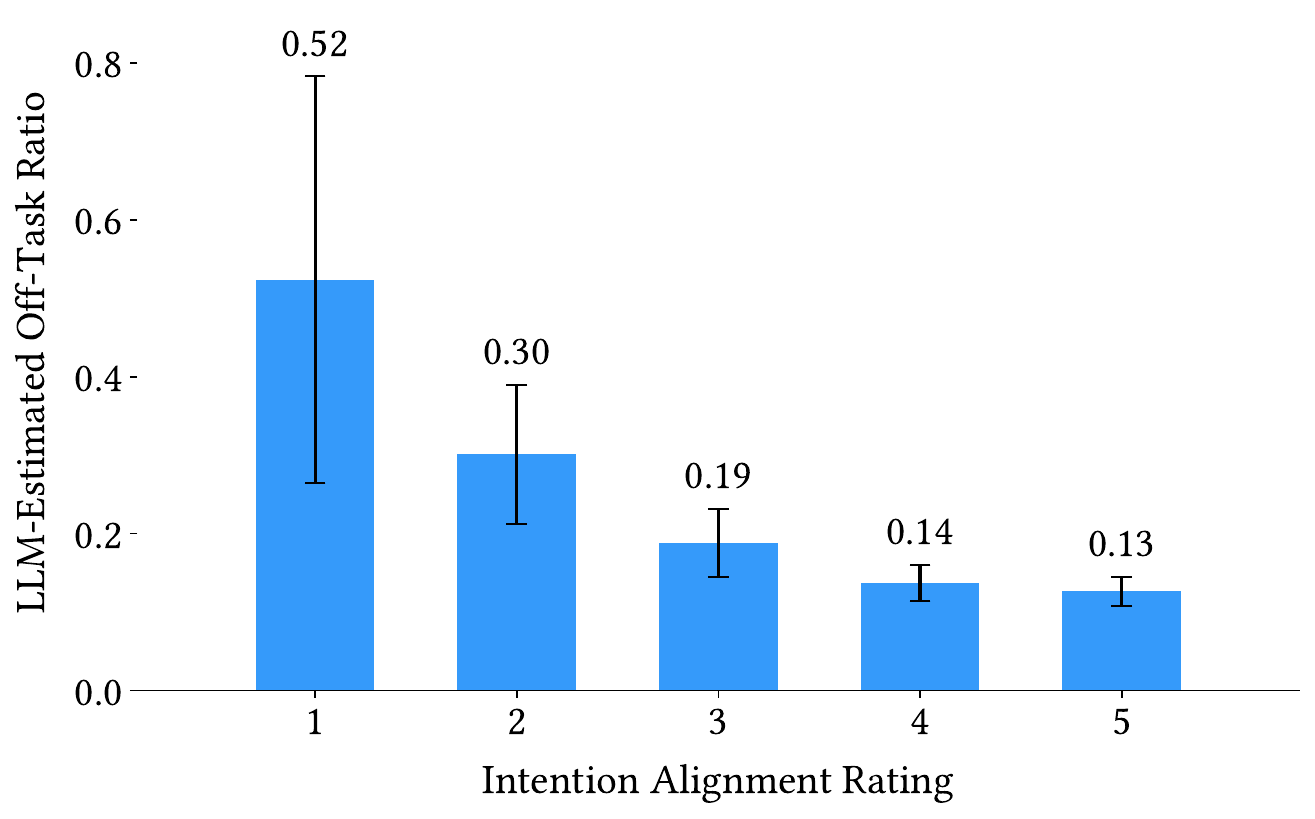}
  \caption{Mean LLM-estimated off-task ratio shown by intention alignment rating (1–5). Error bars denote 95\% confidence intervals across participants (user-averaged).}
  \Description{A bar chart depicting the relationship between intention alignment ratings (1–5) and LLM-estimated off-task ratios.}
  \label{fig:offtask-by-session-alignment}
\end{figure*}

We examined the relation between intention alignment rating and the LLM-estimated off-task ratio, and as shown in Figure~\ref{fig:offtask-by-session-alignment}, higher intention alignment ratings are associated with lower off-task ratios.

\subsection{User Experience Survey Results}\label{app:uxsurvey}
\begin{figure*}[!p]
    \includegraphics[width=\textwidth]{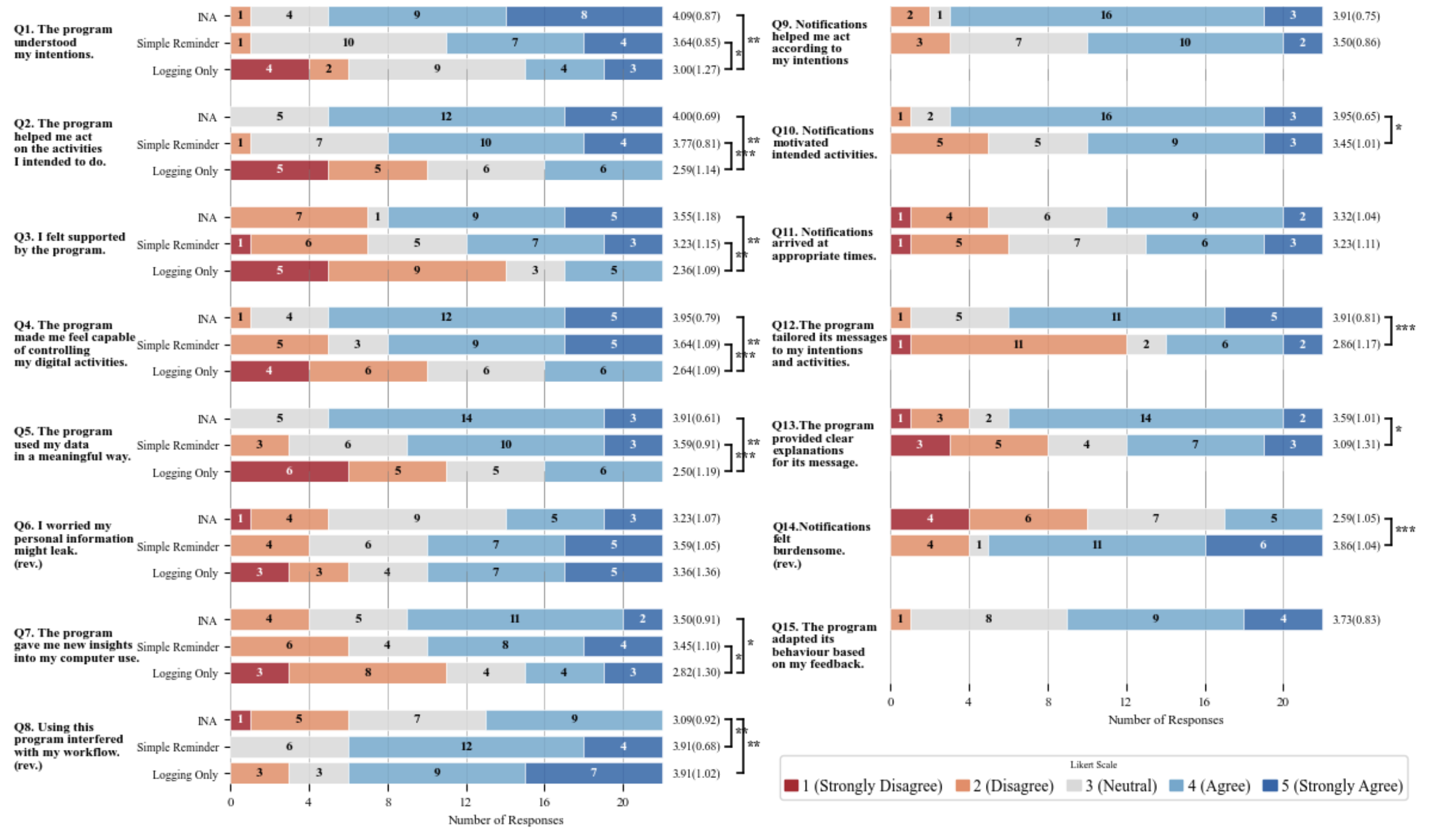}
  \caption{Results of 15-item user experience survey responses by application on a 5-point Likert scale. Means and standard deviations are shown on the right; (rev.) indicates reverse-coded items.  ***, **, and *, indicate significance of p < 0.001, p < 0.01, p < 0.05, respectively.}
  \Description{Bar charts showing user experience survey responses across INA, simple reminder, and logging only system.}
  \label{fig:survey_overall_result}
\end{figure*}

We analyzed the user experience survey by category, dividing items as shown in Table~\ref{tab:survey-items}, where the specific questions for each category can be found. User responses to individual items are presented in Figure~\ref{fig:survey_overall_result}. Items marked with (rev.) are reverse-coded, so that higher values across all items indicate more positive meanings.  
\ours outperformed both \control and \baseline across all dimensions. Participants reported that \ours better understood their intentions, supported them in carrying out intended activities, and delivered notification messages of consistently high quality. However, in Q8 and Q12, some participants also experienced these supportive functions as distracting or burdensome.

\subsection{Interview Coding Reliability Thematic Analysis}\label{app:interview-codes}
A total of 66 interview recordings were transcribed into text, as 22 participants each took part in interviews about three applications. We conducted a codebook thematic analysis that began with inductive coding and iterative refinement. Two coders, including the first author, independently generated initial codes while reading the transcripts and reached consensus on a preliminary codebook. Using this codebook, the coders independently coded the data, compared results, and refined definitions and boundary rules by merging and specifying codes based on observed response patterns. 

To assess intercoder reliability efficiently, the coders independently coded nine transcripts randomly sampled in equal proportions from each application. For each coder, we constructed a binary presence/absence matrix and computed Cohen's~$\kappa$, which indicated substantial agreement ($\kappa = 0.77$). 

Afterward, the two coders each coded the remaining transcripts and cross-checked the results, and finally, the first author reviewed and explained the outcomes to the team. 
The final codebook comprises 24 codes, each with positive, neutral, and negative subcategories, and is summarized in Table~\ref{tab:interview}.

Coding principles were threefold. First, codes were assigned based on the semantic surface of utterances to minimize interpretation. Second, multiple codes could be applied to the same unit. Third, for descriptive statistics, we used respondent-level binary aggregation, such that repeated mentions of the same code by the same participant were counted once.

\subsection{Open-ended Survey Coding Descriptive Analysis}

We applied descriptive coding to analyze the open-ended survey responses. This method assigns short labels to each response segment to summarize the experiences or perceptions reported by participants. The final codebook and results are presented in Table~\ref{tab:coodbook-longtable}.
The results illustrate that \baseline provided unobtrusive logging with little support, \control chadded reminders with mixed sufficiency, and \ours delivered the strongest refocusing effects while introducing new challenges in precision and reliability.

Specifically, \ours frequently reported of ``Return-to-task'' (64.0\% vs.\ 23.0\% \control, 0.0\% \baseline), showing that adaptive prompts often helped participants refocus. 
Mentions of ``Induced immersion'' (32.0\% vs.\ 27.0\% \control, 0.0\% \baseline) further indicate that intention setting and clarification deepened immersion compared to \baseline. 
In the positive codes, \ours uniquely noted ``Real-time notifications'' (36.0\%), while both \ours and \control reported ``Intention declaration effect'' (18.0\%) and similar levels of ``Efficient time use'' (14.0\%) and ``Activity logging \& visualization'' (14.0\%).

Furthermore, negative reports diverged by condition. \ours showed high rates of ``Excessive notifications'' (27.0\%), ``Misclassified notifications'' (27.0\%), and concerns around ``Reliability \& safety'' (27.0\%) and ``UX degradation'' (23.0\%). 
\control more often noted ``Insufficient notifications'' (23.0\%), while \baseline concentrated on ``No observable benefit'' (68.0\%) and the effect-level ``No effect'' (59.0\%). 
Conversely, \baseline participants emphasized ``Non-intrusiveness'' (23.0\%) and ``Visual salience'' (27.0\%), which were rare in \ours (0.0\% and 9.0\%).
\section{Privacy and Security Policies}
\label{app:privacy}

The system was implemented with user privacy and data security as considerations throughout its design. The main policies are as follows.

\paragraph{Data De-identification}
Screen images undergo a de-identification process via a Data Loss Prevention service before storage. The original images used for real-time inference are immediately discarded, and only the masked images, with sensitive information such as names and phone numbers removed, are stored.

\paragraph{Access control}
Access to data storage is strictly controlled based on the principle of least privilege, using multi-factor authentication and Identity and Access Management policies.

\paragraph{Data Encryption}
All data is encrypted in transit between the client and server using the Transport Layer Security protocol. Data at rest is also stored in an encrypted state.

\begin{table*}[!p]
\centering
\resizebox{\textwidth}{!}{%
\begin{tabular}{llp{9.5cm}c}
\toprule
\textbf{Factor} & \textbf{ID} & \textbf{Full question sentence} & \textbf{Cronbach's $\alpha$} \\
\midrule
\multirow{7}{*}{Support}
 & Q1 & The program understood my intentions. & \multirow{7}{*}{0.81} \\
 & Q2 & The program helped me act on the activities I intended to do.  & \\
 & Q3 & I felt supported by the program. & \\
 & Q4 & The program made me feel capable of controlling my digital activities.  & \\
 & Q5 & The program used my data in a meaningful way. & \\
 & Q6 & I worried my personal information might leak. (rev.) & \\
 & Q7 & The program gave me new insights into my computer use. & \\
\midrule
\multirow{1}{*}{Workflow Disruption}
 & Q8 & Using this program interfered with my workflow.  (rev.) & \\
\midrule
\multirow{5}{*}{Message Effectiveness}
 & Q9  & Notifications helped me act according to my intentions. & \multirow{5}{*}{0.78} \\
 & Q10 & Notifications motivated intended activities. & \\
 & Q11 & Notifications arrived at appropriate times. & \\
 & Q12 & The program tailored its messages to my intentions and activities. & \\
 & Q13 & The program provided clear explanations for its message. & \\
\bottomrule
\end{tabular}}
\caption{Survey items grouped by factor. The categorization into \textit{Support}, \textit{Workflow Disruption}, and \textit{Message Effectiveness} was validated by internal consistency, expressed with Cronbach's $\alpha$. Values above 0.70 indicate acceptable reliability, supporting the appropriateness of the factor structure. Cronbach's $\alpha$ was not computed for groups with a single item. Reverse-coded items are marked with (rev.).}
\label{tab:survey-items}
\end{table*}

\begin{table*}[!p]
\centering
\resizebox{\textwidth}{!}{%
\begin{tabular}{llp{8.5cm}ccc}
\toprule
\textbf{Theme} & \textbf{Code} & \textbf{Definition} & \textbf{\baseline (\%)} & \textbf{\control (\%)} & \textbf{\ours (\%)} \\
\midrule
\multirow{5}{*}{Effects}
 & Return-to-task & Notifications interrupt off-task behavior and redirect to intended activity & 0.0 & 23.0 & 64.0 \\
 & Induced immersion & Intention setting ($\rightarrow$ Q\&A interaction) $\rightarrow$ self-evaluation induces deeper immersion & 0.0 & 27.0 & 32.0 \\
 & Deterrence via perceived observation & Perceived observation (system or staff) deters off-task behavior & 14.0 & 9.0 & 9.0 \\
 & Visual salience & System presence / color cues enhance attentional focus & 27.0 & 32.0 & 9.0 \\
 & No effect & No benefit when features feel insufficient/misclassified or hinder use & 59.0 & 14.0 & 9.0 \\
\midrule
\multirow{5}{*}{Pros}
 & Efficient time use & Reduced unnecessary use; more purposeful time allocation & 9.0 & 14.0 & 14.0 \\
 & Intention declaration effect & Entering/refining intentions organizes plans; deepens immersion & 0.0 & 18.0 & 18.0 \\
 & Real-time notifications & Context-appropriate prompts aid return from off-task; convey support & 0.0 & 0.0 & 36.0 \\
 & Activity logging \& visualization & Reviewing what was done and for how long is helpful & 0.0 & 14.0 & 14.0 \\
 & Non-intrusiveness & No obtrusive features; does not interfere with tasks & 23.0 & 0.0 & 0.0 \\
\midrule
\multirow{6}{*}{Cons}
 & No observable benefit & Little/no help; features feel insufficient & 68.0 & 5.0 & 0.0 \\
 & Excessive notifications & Prompts too frequent/nagging; disrupt activity & 0.0 & 9.0 & 27.0 \\
 & Insufficient notifications & Prompts absent or too infrequent; limits effectiveness & 5.0 & 23.0 & 0.0 \\
 & Misclassified notifications & Inaccurate inference leads to context-incongruent messages & 0.0 & 0.0 & 27.0 \\
 & Reliability \& safety concerns & Crashes, lost intentions, privacy concerns & 14.0 & 27.0 & 27.0 \\
 & UX degradation & Impaired workflow (e.g., dual monitors blocked, partial occlusion) & 0.0 & 27.0 & 23.0 \\
\midrule
\multirow{7}{*}{Insights}
 & Frequent social-media checking & Frequent checking of messengers/SNS & 14.0 & 9.0 & 23.0 \\
 & Low sustained attention & Short attention spans; drift within $\sim$20 minutes & 27.0 & 36.0 & 36.0 \\
 & High/improved attention & Rarely straying off-task; reduced off-task proportion & 9.0 & 18.0 & 18.0 \\
 & Notification sensitivity & Stress / heightened reactivity when prompts appear & 0.0 & 5.0 & 5.0 \\
 & Frequent video consumption & Frequent YouTube/OTT video use & 0.0 & 5.0 & 5.0 \\
 & Routine awareness & Awareness of app-specific usage amounts/sequences & 0.0 & 0.0 & 9.0 \\
 & None/unsure & No clear realizations or unsure & 50.0 & 41.0 & 0.0 \\
\bottomrule
\end{tabular}}
\caption{Open-ended survey themes with percentages by condition (\baseline / \control / \ours). Values are the proportion of participants\(N=22\) reporting each code.}
\label{tab:coodbook-longtable}
\end{table*}

\end{document}